\documentclass[10pt,letterpaper]{article}
\usepackage[top=0.85in,left=1.25in,footskip=0.75in]{geometry}

\usepackage{amsmath,amssymb}
\usepackage{changepage}
\usepackage[utf8x]{inputenc}
\usepackage{bbm}
\usepackage{cite}
\usepackage{nameref,hyperref}
\usepackage[right]{lineno} 
\usepackage{microtype}
\DisableLigatures[f]{encoding = *, family = * }

\newlength\savedwidth

\usepackage{setspace} 

\usepackage[aboveskip=1pt,labelfont=bf,labelsep=period,justification=raggedright,singlelinecheck=off]{caption}

\makeatletter
\renewcommand{\@biblabel}[1]{\quad#1.}
\makeatother

\date{}

\usepackage{lastpage,fancyhdr}
\usepackage[]{graphicx} 
\usepackage{epstopdf}
\fancyheadoffset[L]{2.25in}
\fancyfootoffset[L]{2.25in}
\usepackage{placeins}	

\newcommand{\vecu}[1]{\mathbf{#1}}
\newcommand{\matu}[1]{{\pmb #1}}
\newcommand{\angleave}[1]{\left\langle #1 \right\rangle}

\renewcommand{\textmu}{$\mu$}

\usepackage{color}
\newcommand{\mynote}[1]{}
\newcommand{\doi}[1]{\href{http://dx.doi.org/#1}{doi:~#1}}

\mathchardef\mhyphen="2D

\newcommand{\coS}{\mathrm{co}\mhyphen S}
\newcommand{\coP}{\mathrm{co}\mhyphen P}
\newcommand{\coD}{\mathrm{co}\mhyphen D}
\newcommand{\coC}{\mathrm{co}\mhyphen C}
\newcommand{\micron}{\mu\mathrm{m}}

\renewcommand{\a}{\vecu a}
\newcommand{\e}{\vecu e}
\renewcommand{\l}{\vecu l}
\newcommand{\m}{\vecu m}
\newcommand{\n}{\vecu n}
\newcommand{\s}{\vecu s}
\renewcommand{\u}{\vecu u}
\renewcommand{\v}{\vecu v}
\newcommand{\w}{\vecu w}
\newcommand{\x}{\vecu x}

\begin{document}
\vspace*{0.2in}

\begin{flushleft}
{\Large
\textbf{Quantification of Nematic Cell Polarity in Three-dimensional Tissues
}\\[1mm]
}
Andr\'e Scholich\textsuperscript{1},
Simon Syga\textsuperscript{1,2},
Hern\'an Morales-Navarrete\textsuperscript{3},
Fabi\'an Segovia-Miranda\textsuperscript{3},
Hidenori Nonaka\textsuperscript{3},
Kirstin Meyer\textsuperscript{3},
Walter de Back\textsuperscript{2,4},
Lutz Brusch\textsuperscript{2},
Yannis Kalaidzidis\textsuperscript{3},
Marino Zerial\textsuperscript{3,5,6},
Frank J\"ulicher\textsuperscript{1,5,6},
Benjamin M. Friedrich\textsuperscript{5,6,7*}
\\
\bigskip
\textbf{1} Max Planck Institute for the Physics of Complex Systems, Dresden, Germany
\\
\textbf{2} Centre for Information Services and High Performance Computing, TU Dresden, Dresden, Germany
\\
\textbf{3} Max Planck Institute of Molecular Cell Biology and Genetics, Dresden, Germany
\\
\textbf{4} Institute for Medical Informatics and Biometry, Faculty of Medicine Carl Gustav Carus, TU Dresden, Dresden, Germany
\\
\textbf{5} Center for Advancing Electronics Dresden, TU Dresden, Germany
\\
\textbf{6} Cluster of Excellence Physics of Life, TU Dresden, Germany
\\
\textbf{7} Institute for Theoretical Physics, TU Dresden, Germany
\bigskip


* benjamin.m.friedrich@tu-dresden.de
\end{flushleft}

\section*{Abstract} 
How epithelial cells coordinate their polarity to form functional tissues is an open question in cell biology. 
Here, we characterize a unique type of polarity found in liver tissue, nematic cell polarity, 
which is different from vectorial cell polarity in simple, sheet-like epithelia.
We propose a conceptual and algorithmic framework to characterize complex patterns of polarity proteins on the surface of a cell 
in terms of a multipole expansion. 
To rigorously quantify previously observed tissue-level patterns of nematic cell polarity 
(Morales-Navarette et al., eLife 8:e44860, 2019),
we introduce the concept of co-orientational order parameters, 
which generalize the known biaxial order parameters of the theory of liquid crystals. 
Applying these concepts to three-dimensional reconstructions of single cells from 
high-resolution imaging data of mouse liver tissue,
we show that the axes of nematic cell polarity of hepatocytes 
exhibit local coordination and 
are aligned with the biaxially anisotropic sinusoidal network for blood transport.
Our study characterizes liver tissue as a biological example of a biaxial liquid crystal. 
The general methodology developed here could be applied to other tissues or in-vitro organoids.

\section*{Author Summary} 
\nolinenumbers 
Cell polarity enables cells to carry out specific functions.
Cell polarity is characterized by the formation of different plasma membrane domains, 
each with specific composition of proteins, phospholipids and cytoskeletal components. 
In simple epithelial sheets, or tube-like tissues such as kidney, 
epithelial cells are known to display a single apical domain, 
facing a lumenal cavity,
and a single basal domain on the opposite side of the cell, 
facing a basal layer of extracellular matrix. 
This apico-basal polarity defines a vector of cell polarity, which provides a direction of fluid transport, 
e.g.,\ from the basal side of the sheet to the lumen-facing side.
In more complex, three-dimensional epithelial tissues, such as liver tissue 
with its complex network of blood-transporting sinusoids, 
the membrane domains of hepatocyte cells display more intricate patterns, including rings and antipodal pairs of apical membrane.
Here, we develop a mathematical framework to precisely characterize and quantify complex polarity patterns.
Thereby, we reveal ordered patterns of cell polarity that span across a liver lobule.
Our new method builds on physical concepts originally developed for ordered phases of liquid crystals.
It provides a versatile tool to characterize the spatial organization of a complex three-dimensional tissue.

\section*{Introduction}

In multi-cellular organisms, 
almost all tissue cells are spatially asymmetric to serve their function inside their host tissue \cite{bryant_2008}. 
This \textit{cell polarity} can be realized by different kinds of physical anisotropies, 
including cell shape, the structural polarity of their cytoskeleton~\cite{drew_2015}, 
or the protein and lipid composition within the cell membrane~\cite{simons_1985, treyer_musch_2015}. 

Here, we focus on the anisotropic distribution of functional membrane domains on the surface of cells, 
and use the term \textit{cell polarity} specifically for this important case.
A prototypical example is the distribution of polarity-specific apical and basal membrane proteins on the surface of epithelial cells 
\cite{bryant_2008}. 

Among the main functions of epithelial tissues are absorption, filtration, and transport of macromolecules~\cite{bryant_2008}.
Simple epithelial tissues usually cover a body surface or line a body cavity and consist of a one-cell thick layer of cells.
Specifically, apical domains form on the luminal side of the tissue that faces the cavity. 
Apical domains are separated by tight junctions from other membrane domains, such as lateral and basal domains.
Lateral domains provide cell-cell adhesion, 
while basal domains form the interface with the basement membrane and extracellular matrix~\cite{simons_1985, treyer_musch_2015}.
This structural asymmetry of apical and basal domains in simple epithelia defines a vectorial cell polarity 
(sometimes referred to as \textit{columnar polarity} \cite{treyer_musch_2015}), see Fig.~\ref{fig:vectorial_and_nematic_cell_polarity}A.
This vectorial cell polarity sets a direction for the \textit{directed transport} of macromolecules.

However, there are also epithelial tissues with a more complex, three-dimensional architecture, 
such as liver tissue \cite{simons_1985,treyer_musch_2015,Kmiec2001,Stamatoglou_1994,decaens_1996}.
The functional unit of the liver, the liver lobule, 
is organized around a central and a portal vein,
which are connected by a dense, three-dimensional network of sinusoids that transport blood 
(see also Fig.~\ref{fig:apical_sinusoid_coop}A).
Hepatocytes, the main cell type of the liver, are evenly distributed in the lobule
with a volume fraction of approximately 80\%~\cite{Kmiec2001}.
Each hepatocyte is in contact with the sinusoidal network at multiple basal membrane domains,
which facilitate the exchange of metabolites with the blood stream~\cite{treyer_musch_2015}.
The sinusoidal network was proposed to provide orientational cues to hepatocytes \cite{Hoehme2010,Sakaguchi2008}.
In addition to the basal contacts,
each hepatocyte possesses multiple apical membrane domains that form narrow lumina with adjacent cells, 
into which bile is excreted~\cite{treyer_musch_2015,Elias1955,Meyer2017}.
These lumina form a second, three-dimensional network, the bile canaliculi network. 
The direction of bile excretion by individual hepatocytes and, 
correspondingly, the distribution of apical membrane domains on their surface,
cannot be characterized by a single vector, yet is also not random.

Previously, Elias put forward an idealized description of liver tissue 
in terms of a crystal-like organization of sinusoids and polarized hepatocytes \cite{elias_1949, elias_1952}.
This model has recently been revisited using high-resolution imaging data of mouse liver tissue \cite{morales-navarette_2015}.
This study showed that the structure of liver tissue is intermediate between an amorphous structure and a perfect crystal, 
best described as a liquid crystal with orientational but not positional order.
The quantification of cell polarity in three-dimensional reconstructions of such high-resolution data
prompts new analysis methods to infer the coordination of cell polarity at the tissue level.

We will characterize the distribution of apical membrane domains on the surface of hepatocytes by a tripod of nematic axes. 
Intuitively, a nematic axis can be thought of as a double-headed arrow that specifies an axis, 
but does not single out any of the two directions parallel to that axis, see Fig.~\ref{fig:vectorial_and_nematic_cell_polarity}B.
We will refer to this characterization as \textit{nematic cell polarity} 
to highlight the analogy to \textit{vectorial cell polarity}
(although a set of nematic axes is not polar in the strict mathematical sense).
 
A first approach was restricted to the analysis of a single type of cell polarity axes at a time \cite{Morales2018}.
Here, we extend the analysis in \cite{Morales2018} to the biaxial case of 
a full tripod of nematic cell polarity axes.
We present a systematic and versatile method to characterize cell polarity by means of a multipole expansion. 
The zeroth moment of this expansion describes a uniform surface density of a polarity marker, 
as found, e.g., in non-polarized mesenchymal cells. 
The first moment of this expansion describes vectorial polarity, 
and characterizes, e.g., apico-basal polarity of cells in simple epithelial sheets.
The second moment defines nematic cell polarity, 
and characterizes, e.g., the more complex distribution of apical membrane domains found in hepatocytes.

We apply the concept of nematic cell polarity to apical membrane patterns of hepatocytes.
We find that the nematic cell polarity of hepatocytes is aligned along curved director fields within the liver lobule, 
in line with previous observations \cite{Morales2018}.
We demonstrate that the coordination of cell polarity is biaxial, 
i.e., its description requires two local reference axes.
Additionally, we find that nematic cell polarity of hepatocytes is 
correlated with the local biaxial anisotropy of the sinusoidal network.
A minimal interaction model conceptualizes the co-alignment of hepatocyte cell polarity and 
the local anisotropy of the sinusoidal network.

The co-orientational order parameters (COOP) introduced here 
to characterize the structure of three-dimensional liver tissue 
naturally generalize previous work on order in effectively two-dimensional cells and tissues. 
Drew et al.\ introduced COOPs for two-dimensional systems, 
and applied this analytical metric to quantify the co-alignment of cytoskeletal structures in muscle cells \cite{drew_2015}.
Other authors addressed planar cell polarity~\cite{Marcinkevicius2009,Sagner2012},
or nematic alignment of cell shape elongation \cite{Saw2017}. 
The COOP introduced here provide a unified framework to characterize such cellular anisotropies also in three space dimensions.

\section*{Results}

\subsection*{Nematic cell polarity}

We present a method to classify distributions of polarity membrane domains
on the surface of cells by a multipole expansion in terms of their spherical power spectrum.
Using this spherical power spectrum, 
we describe the dominant symmetry of such a distribution of membrane proteins in terms of either predominantly vectorial, nematic or higher-order type.
We first illustrate the method using distributions on a sphere, 
and afterwards show how surface distributions on cells of non-spherical shape can be mapped to this case.
For the convenience of the reader, a list of mathematical symbols can be found in SI text \nameref{sec:symbols}.

\begin{figure}[ht] 
	\centering  
    \makebox[0.7\textwidth][r]{
        \includegraphics[width=0.77\textwidth]{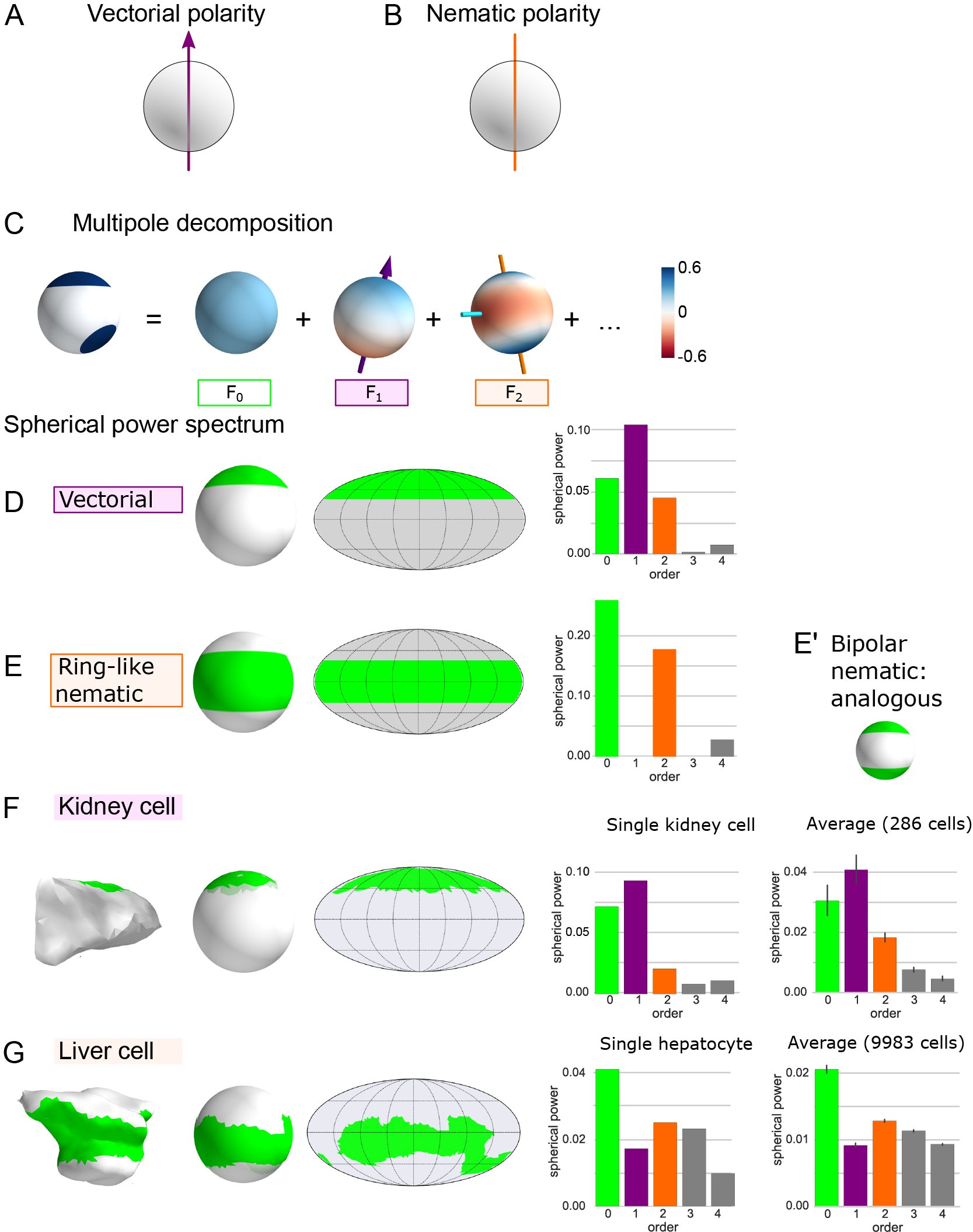}
    }%
	\caption{{\bf Multipole decomposition of surface patterns.}\newline
		(A, B) Schematic of vectorial and nematic cell polarity.
		(C) Multipole decomposition of a distribution on a sphere into spherical harmonics.
		(D) Prototypical membrane distribution of vectorial polarity type with respective Mollweide projection and spherical power spectrum.
		(E) Same as panel D but for a ring-like surface distribution. Here, the second mode of the spherical power spectrum dominates.
		(E') Analogously, for a bipolar surface distribution, the second mode of the spherical power spectrum also dominates (not shown).
		(F) Spherical projection, Mollweide projection and spherical power spectrum for an epithelial tubular cell from kidney tissue, 
		as well as averaged power spectrum for a population of cells ($n=286$).	
		(G) Same as panel F, but for a hepatocyte from mouse liver tissue,
as well as a population of hepatocytes ($n=9983$).}
	\label{fig:vectorial_and_nematic_cell_polarity}
\end{figure} 

Let $f(\x)$ with $\x\in \mathcal{S}^2$ represent an area density on the surface of the unit sphere $\mathcal{S}^2$.
Similar to the two-dimensional Fourier transform for functions defined on a plane, 
we decompose the density $f(\x)$ into orthogonal modes
\begin{align}
\label{eq:Y}
 f(\x) 
   = \sum\limits_{l=0}^{\infty} F_{l}(\x)
   \quad .
\end{align}
Here, the mode $F_l(\x)$ of degree $l$ is given by $F_{l}(\x) = \sum_{m=-l}^{l} f_{l}^{m} Y_{l}^{m}(\x)$, 
where 
$Y_{l}^{m}(\x)$ denotes the spherical harmonic of degree $l$ and order $m$ (normalized to unity).
%
Using the ortho-normality of the spherical harmonics, the expansion coefficients $f_{l}^{m}$ are given by 
$f_{l}^{m} = \int_{\mathcal{S}^2} \mathrm{d}^2\x \, f(\x) \, Y_l^{m*}(\x) $.
Here, integration is performed over the unit sphere $\mathcal{S}^{2}$ (with respect to the standard Euclidean measure);
the star denotes the complex conjugate.
A visual representation of this spherical decomposition is given in Fig.~\ref{fig:vectorial_and_nematic_cell_polarity}C:
the zeroth mode $F_0$ is isotropic and encodes the mean of the surface distribution $f(\x)$.
The first mode $F_1(\x)$ can be represented by a vector 
that points to the spherical average of the surface distribution\cite{fisher_1953}. 
The second mode $F_2(\x)$ is related to nematic polarity and will be at the focus of this work.
The possible existence of higher modes is indicated.
The original distribution can be restored by summing up all modes.

In analogy to Fourier analysis of one-dimensional signals, we define the power $S_{f\!f}(l)$ of each spherical mode $F_l(\x)$ 
as its $L^2$-norm (normalized by the surface area of the unit sphere)  
\begin{align}
S_{f\!f}(l) = \frac{1}{4\, \pi} \int_{\mathcal{S}^2} \mathrm{d}^2\x\, | F_{l} |^{2} 
\quad.
\end{align}
This defines the \textit{spherical power spectrum}, 
for which a generalized Parseval's theorem holds,
$\sum_l S_{f\!f}(l)=\int \mathrm{d}^2\x\,|f(\x)|^2$.

Fig.~\ref{fig:vectorial_and_nematic_cell_polarity}D and Fig.~\ref{fig:vectorial_and_nematic_cell_polarity}E show prototypical vectorial and nematic distributions and their respective spherical power spectra.
We visualize surface patterns also as Mollweide projections, 
an equal-area, pseudocylindrical geographic projection \cite{snyder_1987}.
The spherical power spectrum of the cap-like distribution, shown in Fig.~\ref{fig:vectorial_and_nematic_cell_polarity}D, 
has a clear peak at the first mode, corresponding to a predominantly vectorial polarity type of the surface distribution.
In contrast, for a ring-like pattern as shown in Fig.~\ref{fig:vectorial_and_nematic_cell_polarity}E
(and, analogously, for a bipolar pattern with two antipodal caps, see Fig.~\ref{fig:vectorial_and_nematic_cell_polarity}E'), 
all odd modes of the spherical power spectrum, including the first mode, 
vanish by symmetry. 
The power spectrum attains its maximum at the second mode, 
which classifies these distributions as nematic.

Biological cells are not perfectly spherical.
We propose a simple method to project distributions on the surface of star-convex shapes onto a sphere 
(i.e., we require that shape can be represented as distance from a common center, which is taken to be the origin, as a single-valued function of the solid angle).
This method allows to analyze the anisotropy of surface patterns independent of any anisotropy of cell shape.
For hepatocytes, the correlation between cell shape and apico-basal cell polarity is weak, for a discussion see \cite{Morales2018}.

Specifically, we radially project from the star-convex shape to a unit sphere concentric with the shape, 
retaining the nominal value of the original distribution,  
see SI text \nameref{sec:spherical_projection} for additional details.
To each cell with surface distribution $\rho(\vecu x)$ of (apical) polarity proteins, 
we associate the projection $f(\x)$ of this distribution on the unit sphere $\mathcal{S}^2$.
Examples of this projection for an epithelial tubular cell from kidney tissue and 
a hepatocyte from liver tissue are shown in Fig.~\ref{fig:vectorial_and_nematic_cell_polarity}F 
and Fig.~\ref{fig:vectorial_and_nematic_cell_polarity}G, respectively.
Apical plasma membrane domains of these cells from kidney and liver tissue, respectively, were identified by staining fixed cells 
with anti-CD13 (Novus, cat NB100-64843, rat, 1/500) as reported previously \cite{morales-navarette_2015,Morales2018}.
The kidney cell exhibits clear vectorial polarity as reflected by a peak of the spherical power spectrum at the first mode.
This is expected as kidney cells are regarded to belong to the vectorial cell polarity type also present in sheet-like epithelia~\cite{treyer_musch_2015}.
This observation from a typical kidney cell is confirmed for an ensemble of cells ($n=286$);
note that the relative magnitudes of spherical power modes exhibit a consistent pattern, 
whereas their absolute magnitude scales with the square of the projected area fraction of apical membrane.

In contrast, for the hepatocyte, we find a dominant second mode, while the first mode is less pronounced.
If spherical power spectra are averaged over a population of cells, 
we still find a dominant first mode for the case of kidney cells, see Fig.~\ref{fig:vectorial_and_nematic_cell_polarity}F,
and a pronounced second mode that exceeds the first mode in the case of hepatocytes, 
see Fig.~\ref{fig:vectorial_and_nematic_cell_polarity}G.
Note that the large ensemble of cells analyzed here may contain segmentation errors;
thus the ensemble-averaged spherical power spectra presented in Fig.~\ref{fig:vectorial_and_nematic_cell_polarity}FG
represent a lower bound.

This analysis of spherical power spectra 
highlights the structural difference between these two different cell types and prompts for a description of hepatic cell polarity in terms of \textit{nematic cell polarity}.
We introduce the nematic tensor $\matu{A}$ of the spherical distribution $f(\x)$, 
\begin{align}
\matu{A} = \frac{1}{2} \int_{\mathcal{S}^2} \! \mathrm{d}^2\x \, f(\x)\left( 3\, \x\otimes\x - {\mathbbm 1}\right)
\quad ,
\label{eq:nematic_tensor}
\end{align}
where $\mathbbm{1}$ denotes the identity tensor with components $\mathbbm{1}_{\alpha\beta}=\delta_{\alpha\beta}$.
The nematic tensor $\matu{A}$ encodes the same information as the second multipole $F_2(\x)$.
More generally, there is a formal link between the spherical modes of order $l$ 
and the reduced Cartesian multipole moments \cite{rosso_2007}, 
see also SI text \nameref{sec:relation_second_mode}.
The nematic tensor $\matu{A}$ is closely related to a moments-of-inertia tensor, see SI test \nameref{sec:cuboid_visualization}.
We order the eigenvalues $\alpha_1$, $\alpha_2$, $\alpha_3$ of $\matu{A}$ 
such that $\alpha_1 \geq \alpha_3 \geq \alpha_2$ holds
and denote the eigenvectors corresponding to $\alpha_1$, $\alpha_2$, $\alpha_3$
by $\a_1$, $\a_2$, $\a_3$
Motivated by Fig.~\ref{fig:inertia_box}AB, 
we will refer to $\a_2$ as the \textit{ring axis} 
and $\a_1$ as the \textit{bipolar axis}.
Below, the axis $\a_2$ will represent an example of a \textit{first} principal axis 
used to define co-orientational order parameters,
while the axis $\a_1$ will represent the \textit{second} principal axis.
The numbering of axes $\a_2$ and $\a_1$ was chosen to be consistent with \cite{Morales2018} (there $\alpha_i=\sigma_i$, $i=1,2,3$).

\subsection*{Cuboid representation of nematic cell polarity}

\begin{figure}[!b] 
\centering
\includegraphics[width=0.7\textwidth]{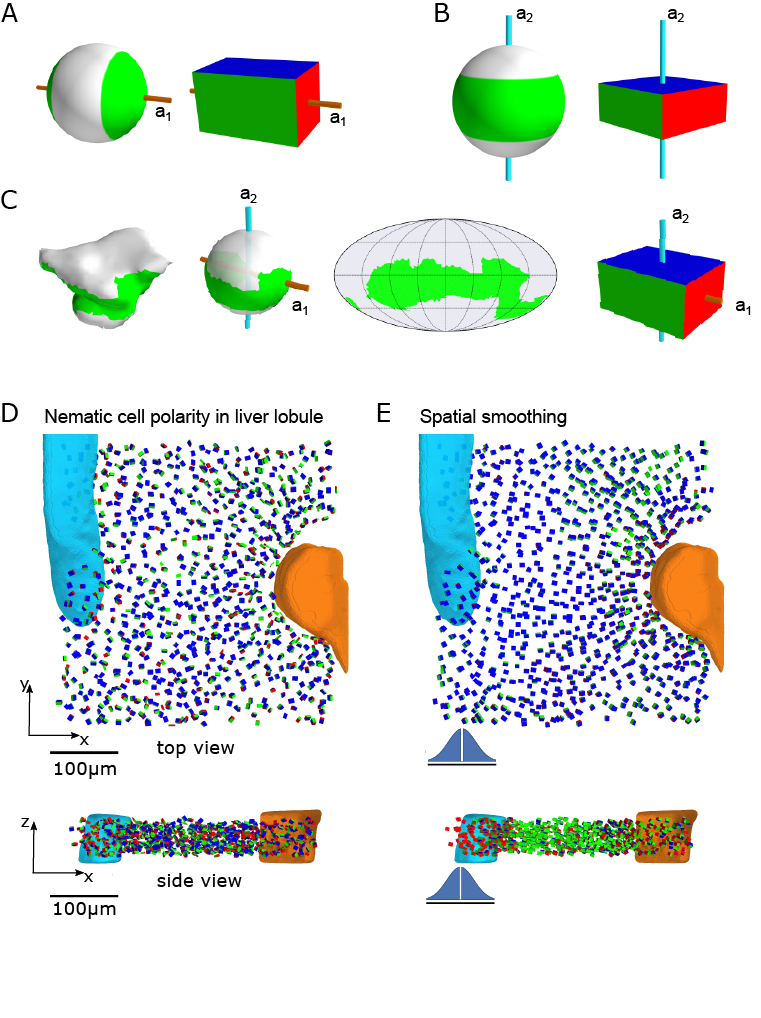} 
\caption{
{\bf Spatial patterns of nematic cell polarity.}
\newline
We visualize surface distributions by cuboids that have the same moments of inertia tensor.
Opposite faces of these cuboids are colored red, green, and blue, respectively,
corresponding to the principal axes of inertia (ordered in increasing order).
(A)
Idealized bipolar distribution. 
The bipolar axis $\a_1$ (golden, corresponding to smallest moment of inertia)
determines the position of the red faces.
(B)
Idealized ring-like distribution. 
The ring axis $\a_2$ (cyan, corresponding to largest moment of inertia)
determines the position of the blue faces.
(C) 
Apical membrane distribution for a typical hepatocyte, 
spherical projection, Mollweide projection, and equivalent cuboid 
with two distinguished principal axes of inertia $\a_1$ and $\a_2$,
corresponding to the bipolar and ring nematic cell polarity axes, respectively. 
(D) 
For each hepatocyte in a tissue sample, 
the corresponding cuboid is plotted, 
revealing ordered patterns at the liver lobule level. 
(E) 
Orientational order becomes even more apparent after spatial averaging, 
which was performed using a Gaussian kernel with standard deviation of 20\,\textmu m and omitting the cell in the center
(kernel sketched to scale, blue), 
see SI text \nameref{sec:gaussian_average_nematic_tensors} for details.
In panels (D) and (E), a central vein (cyan) and a portal vein (orange) are shown, which serve as landmarks within a liver lobule.
}
\label{fig:inertia_box}
\end{figure}

To qualitatively assess putative spatial patterns of nematic cell polarity, 
we propose a visualization method in terms of equivalent cuboids,
see Fig.~\ref{fig:inertia_box}A-C.
Mathematically, the cuboid for a cell
is uniquely determined by the condition that its traceless moments-of-inertia tensor should equal 
the traceless moments-of-inertia tensor of the spherical distribution $f(\x)$, 
see SI text \nameref{sec:cuboid_visualization} for details.
Briefly, the edges of the cuboid are parallel to the eigenvectors of the nematic tensor $\matu{A}$ 
associated to $f(\x)$,
while the side-lengths of the cuboid depend on the eigenvalues of $\matu{A}$.

Fig.~\ref{fig:inertia_box}A shows an idealized bipolar distribution and its equivalent cuboid.
Here, the longest edge of the cuboid is parallel to the bipolar axis $\a_1$ of the surface distribution, 
while the two shorter axes have equal length.
Similarly, for an idealized ring-like distribution, 
the shortest edge of the cuboid is parallel to the ring axis $\a_2$ of the surface distribution, 
while the two longest edges have equal length, 
see Fig.~\ref{fig:inertia_box}B.
We colored opposite faces of the cuboids in red, green, and blue, 
where red corresponds to the bipolar axis $\a_1$, and 
blue to the ring axis $\a_2$.
Fig.~\ref{fig:inertia_box}C shows the cuboid representation of a typical hepatocyte (with apical membrane distribution shown in green).

Using this cuboidal representation,
we can visualize biaxial nematic cell polarity with respect to apical membrane distribution of all hepatocytes within a tissue section, 
see Fig.~\ref{fig:inertia_box}D.
There, part of a liver lobule is shown with characteristic landmarks represented by 
the portal vein (orange) and the central vein (cyan).
In top view, 
most of the polarity cuboids are faced with their blue side up
(indicating an approximately parallel alignment of the ring axis $\a_2$ 
of hepatocyte polarity with the large veins; for the chosen tissue sample these are approximately parallel to the $z$ axis).
This highlights the existence of a lobule-wide pattern of spatial order.
The tissue-level alignment of nematic cell polarity becomes even more apparent 
when polarity fields are locally averaged to reduce fluctuations, 
see Fig.~\ref{fig:inertia_box}D.
Next, we will use order parameters from the theory of liquid crystals 
to quantify the observed spatial patterns of aligned cell polarity.

\subsection*{Order parameters of nematic cell polarity}
\label{sec:SPDC}

We can quantify orientational order of nematic cell polarity within a tissue in terms of orientational order parameters (OOP), $S$, $P$, $D$, $C$.
These order parameters were originally developed for the study of biaxial order in liquid crystals \cite{straley_1974, luckhurst_2015}. 
We briefly review their definition and provide illustrative examples to convey their geometric meaning.

Before we present the formal definition of $S$, $P$, $D$, $C$, we want to motivate the different roles played by the four OOPs.
For an ensemble of uniaxial objects, each characterized by a single principal axis,
$S$ quantifies how well the ensemble of these principal axes are aligned to a common mean direction. 
The mean direction of this nematic alignment defines one of several reference axes introduced below. 
In the theory of simple liquid crystals, 
$S$ is the most widely used OOP (and probably the most important one).
Like $S$, the OOP $P$ is also defined already for an ensemble of uniaxial objects (with a single principal axis each).
While $S$ characterizes the strength of nematic alignment of the single principal axis in the ensemble, 
$P$ characterizes anisotropic fluctuations of this principal axis, 
i.e., $P$ becomes non-zero if the deviations of the principal axes from the mean direction are skewed in a particular direction.
The direction of anisotropic fluctuations defines a second reference axis.
The two other OOPs, $D$ and $C$, are only defined for biaxial objects that are characterized by \textit{two} principal axes (a third principal axis can be deduced from the two). 
The OOP $C$ characterizes nematic alignment of the second principal axis towards a second reference axis. 
Finally, $D$ quantifies an `unexpected relation' between the fluctuations of the first and second principal axis, respectively, 
with respect to their alignment to a first reference axis of mean alignment. 
For a maximum-entropy distribution of orientations, constrained to specific numerical values of $S$, $P$, $C$, the OOP $D$ would be zero.

In liquid crystals, built up by an ensemble of anisotropic molecules, 
a non-zero value of $S$ can arise from the interactions of uniaxial molecules (e.g., molecules approximated as a rod) with a uniaxial external field. 
Here, the axis of the external field sets the direction of mean alignment, i.e., the first reference axis. 
A non-zero value of $P$, however, additionally requires either a second external field, orthogonal to the first one, 
or boundary conditions that break rotational symmetry for rotations around the first reference axis. 
Non-zero values of $D$ and $C$ obviously require objects that are intrinsically biaxial, i.e., that do not possess rotational symmetry around their first principal axis. 
While a non-zero value of $D$ can result already from interactions of biaxial objects with an external uniaxial field, 
a non-zero value of $C$ requires either an external field that is biaxial, boundary conditions that break uniaxial symmetry, or, possibly, biaxial inter-molecular interactions. 

We now proceed to the formal definition of OOPs, which will later be generalized to the case of co-orientational order parameters and applied to quantitatively characterize design principles of liver tissue. 
We consider an ensemble of cuboids, 
each characterized by a tripod of principal axes, which we characterize by orthonormal vectors $\vecu n$, $\vecu m$, $\vecu l$.
These unit vectors are only defined up to sign, 
thus any meaningful physical quantity should be invariant under sign flips $\n\rightarrow -\n$, $\m\rightarrow -\m$, $\l\rightarrow -\l$.
Formally, each cuboid is said to possess so-called $D_{2h}$-symmetry, i.e., 
it is invariant under line reflections at its principal axes.
We distinguish
a \textit{first principal axis} $\vecu n$, as well as 
a \textit{second principal axis} $\vecu{m}$, and
a \textit{third principal axis} $\vecu l$.
It is convenient to introduce, for each cuboid, 
two traceless tensors $\matu Q$ and $\matu B$ that characterize its tripod of axes \cite{matteis_2008, luckhurst_2015}
\begin{equation}
\label{eq:QB}
 \matu Q = \frac{1}{2} \left( 3 \, \vecu n \otimes \vecu n - \mathbbm{1}\right) \quad , \quad
 \matu B = \frac{3}{2} \left( 
\vecu l \otimes \vecu l - \vecu m \otimes \vecu m 
\right) \quad .
\end{equation}
Here, $\mathbbm{1}$ is the identity tensor and $\otimes$ denotes the outer product.

If the principal axes $\n$, $\m$, $\l$ of each tripod 
are given by the normalized eigenvectors of a nematic tensor $\matu{A}$ 
(e.g., the nematic tensor of a projected surface distribution $f(\x)$ of membrane proteins), 
we can recover $\matu{A}$ as linear superposition of the two tensors $\matu{Q}$ and $\matu{B}$, 
see SI text \nameref{sec:relation_biaxial_order_invariants}.
The mathematical advantage of the traceless tensors $\matu Q$ and $\matu B$ is that they conveniently allow to compute ensemble averages, $\langle \matu Q \rangle$ and $\langle \matu B \rangle$.
The eigenvalues of these averaged tensors provide important invariants of orientational order
\begin{align}
\label{eq:rq}
\matu{R}_Q^T \, \langle \matu{Q} \rangle \, \matu{R}_Q 
&= 
\left( 
\begin{array}{ccc}
                     -\frac{1}{2} (S-P) & 0 & 0 \\
                     0 &-\frac{1}{2} (S+P) & 0 \\
                     0 & 0 & S\\
\end{array} \right)
\quad , 
\\
\label{eq:rb}
\matu{R}_B^T \, \langle \matu{B} \rangle \, \matu R_B 
&= 
\left( 
\begin{array}{ccc}
                     -\frac{1}{2} (D-3\,C) & 0 & 0 \\
                     0 &-\frac{1}{2} (D+3\,C) & 0 \\
                     0 & 0 & D\\                    
\end{array} 
\right) 
\quad .
%
\end{align}
Here, $\matu R_Q$ and $\matu R_B$ are rotation matrices 
that diagonalize $\angleave{\matu Q}$ and $\angleave{\matu B}$, respectively.

In principle, the rotation matrices $\matu{R}_Q$ and $\matu{R}_B$ might be different.
However, for important special cases, e.g., ensembles of biaxial molecules interacting with simple external fields,
both rotation matrices can be chosen equal, $\matu{R}=\matu{R}_Q=\matu{R}_B$.
In this case, the ensemble-averaged tensors $\langle\matu{Q}\rangle$ and $\langle\matu{B}\rangle$ possess the same eigenvector basis,
given by three mutually orthogonal unit vectors $\w$, $\v$, $\u$.
The physical meaning of $\w$, $\v$, $\u$ is that these vectors define  
mutually orthogonal symmetry axes, such that the statistics of the ensemble of cuboids is invariant under line reflections at these axes.
The ensemble of cuboids is said to possess \textit{$D_{2h}$-symmetry} in this case.
In the case of $D_{2h}$-symmetry, 
the common eigenvectors $\w$, $\v$, $\u$ of $\angleave{\matu Q}$ and $\angleave{\matu B}$ define 
a \textit{director frame of reference axes} of the ensemble.
Note $\matu{R}_Q=\matu{R}_B = [\u,\v,\w]^T$ (with row-vectors $\u^T$, $\v^T$, $\w^T$).

Using Eq.~\eqref{eq:rq} and Eq.~\eqref{eq:rb},
we can rewrite $S$, $P$, $D$, $C$ as averaged direction cosines~\cite{luckhurst_2015}
\begin{align}
\label{eq:SPDC}
S &= \frac{1}{2} \left\langle 3 (\n^{(i)}\cdot\w)^2 - 1             \right\rangle \quad, \\
P &= \frac{3}{2} \left\langle   (\n^{(i)}\cdot\u)^2 - (\n^{(i)}\cdot\v)^2 \right\rangle \quad, \notag \\
D &= \frac{3}{2} \left\langle  (\l^{(i)}\cdot\w)^2 - (\m^{(i)}\cdot\w)^2  \right\rangle \quad, \notag \\
C &= \frac{1}{2} \left\langle 
	(\l^{(i)}\cdot\u)^2 - (\l^{(i)}\cdot\v)^2 +
	(\m^{(i)}\cdot\v)^2 - (\m^{(i)}\cdot\u)^2 
                                \right\rangle \quad , \notag
%
%
\end{align}
where $\langle\cdot\rangle$ averages over an ensemble 
of nematic axes $\n^{(i)}$, $\m^{(i)}$, $\l^{(i)}$ indexed by $i$, 
using a fixed tripod of reference axes $\w$, $\v$, $\u$. 

In Eq.~(\ref{eq:SPDC}), there is ambiguity regarding the ordering of the various axes, 
$\n$, $\m$, $\l$, and $\w$, $\v$, $\u$.
For our specific case, 
the mapping from the nematic cell polarity axes $\vecu a_1$, $\vecu a_2$, $\vecu a_3$ to
the principal axes $\vecu n, \vecu m, \vecu l$ is only defined up to a constant permutation
\begin{equation}
\n=\a_{\pi(2)} \quad,\quad 
\m=\a_{\pi(1)} \quad,\quad 
\l=\a_{\pi(3)}, \quad \pi\in S_3 \quad ,
\end{equation}
where $S_3$ denotes the group of all permutations of the indices $(1,2,3)$.
Note that the same permutation $\pi$ must be used for all tripods of nematic axes 
$\a_1^{(i)}$, $\a_2^{(i)}$, $\a_3^{(i)}$ of the ensemble.
Similarly, 
if $\e_1$, $\e_2$, $\e_3$ denote the common eigenvectors of 
$\langle\matu Q\rangle$ and $\langle\matu B\rangle$ 
(with some fixed ordering),
we distinguish
a \textit{first reference axis} $\w=\e_{\rho(2)}$ 
from a \textit{second reference axis} $\v=\e_{\rho(1)}$,
and a \textit{third reference axis} $\u=\e_{\rho(3)}$,
where $\rho\in S_3$ denotes a permutation of reference axes.
(The axes $\w$, $\v$, $\u$ are also called director axes~\cite{luckhurst_2015}.)

We distinguish two different choices for $\pi$ and $\rho$, which give rise to 
orientational order parameters (OOP) as commonly used in the theory of liquid crystals \cite{luckhurst_2015}, and
co-orientational order parameters (COOP) introduced here.

A common choice, put forward, e.g., by Zannoni et al.~\cite{luckhurst_2015} in the field of biaxial nematics, 
is to chose the permutations $\pi$ and $\rho$ of principal and reference axes such that 
\begin{equation}
|S| \text{ is maximal, } P\ge 0, \text{ and } C\ge 0.
\label{eq:zannoni}
\end{equation}
This condition specifies an ordering of both principal and reference axes.
The tensor $\angleave{\matu Q}$ and the scalar order parameters $S$ and $P$ 
quantify alignment of the first principal axis $\n$, 
whereas the tensor $\angleave{\matu B}$ and the scalar order parameters $D$ and $C$ 
quantify the alignment of the second and third principal axes, $\m$ and $\l$.
We will refer to the values of $S$, $P$, $D$, $C$ 
corresponding to the ordering of nematic axes specified by Eq.~(\ref{eq:zannoni}) as
\textit{order parameters} (OOP) without further specification.
Note that different normalization conventions for OOP are in use, an overview can be found in \cite{rosso_2007}.
While OOPs are always well-defined, 
they have a crucial disadvantage: 
OOPs may change discontinuously if system parameters are smoothly varied due to abrupt changes of either $\pi$ or $\rho$, 
see SI text \nameref{sec:relation_biaxial_order_invariants}.

We propose an alternative choice, where the ordering of axes is directly determined by 
the properties of a nematic tensor $\matu A$.
In the case of surface distributions on a sphere considered here,
we take $\vecu n$ to point in the direction of the ring axis $\a_2$, and 
$\vecu m$ to point in the direction of the bipolar axis $\a_1$.

We consider the general case, 
where for each nematic tensor $\matu{A}^{(i)}$ from an ensemble of tensors indexed by $i$, 
we additionally have a second nematic tensor $\matu{E}^{(i)}$ for each $i$.
Below, we discuss two natural cases of such reference tensors.
Let $\varepsilon_1$, $\varepsilon_2$, $\varepsilon_3$ be the eigenvalues of one of the $\matu{E}^{(i)}$,
ordered such that $\varepsilon_1\ge\varepsilon_3\ge\varepsilon_2$,
and $\e_1$, $\e_2$, $\e_3$ be the corresponding (normalized) eigenvectors.
We introduce a tripod of reference axes for each index $i$ as 
$\w^{(i)}=\e_2$, $\v^{(i)}=\e_1$, $\u^{(i)}=\e_3$,
i.e., we chose $\rho$ as the identify permutation, $\rho=\mathrm{id}$.
We define \textit{co-orientational order parameters} (COOP) by generalizing Eq.~(\ref{eq:SPDC})
to this general case, where the reference axes $\u^{(i)}$, $\v^{(i)}$, $\w^{(i)}$ are derived from a set of reference tensors $\matu{E}^{(i)}$
\begin{align}
\label{eq:coSPDC}
\coS &= \frac{1}{2} \left\langle 3 (\n^{(i)}\cdot\w^{(i)})^2 - 1                         \right\rangle \quad, \\
\coP &= \frac{3}{2} \left\langle (\n^{(i)}\cdot\u^{(i)})^2 - (\n^{(i)}\cdot\v^{(i)})^2   \right\rangle \quad, \notag \\
\coD &= \frac{3}{2} \left\langle   (\l^{(i)}\cdot\w^{(i)})^2 - (\m^{(i)}\cdot\w^{(i)})^2 \right\rangle \quad, \notag \\
\coC &= \frac{1}{2} \left\langle   
		  (\l^{(i)}\cdot\u^{(i)})^2 - (\l^{(i)}\cdot\v^{(i)})^2 
		+ (\m^{(i)}\cdot\v^{(i)})^2 - (\m^{(i)}\cdot\u^{(i)})^2         
                    \right\rangle \quad , \notag
\end{align}
where $\langle\cdot\rangle$ averages over the ensemble of pairs of tripods indexed by $i$.
We propose a scheme to compute reference tensors $\matu{E}^{(i)}$ for the important case, 
where the tensors $\matu{A}^{(i)}=\matu{A}(\x^{(i)})$ depend on spatial position $\x^{(i)}$.
For each position $\x^{(i)}$, we define $\matu{E}^{(i)}=\langle \matu{A}(\x^{(i)}) \rangle_\mathrm{loc}$ 
using a local average with a ``punctured'' three-dimensional Gaussian kernel centered at $\x^{(i)}$ 
(excluding the tensor $\matu{A}^{(i)}$ at the central position $\x^{(i)}$), 
see SI text \nameref{sec:gaussian_average_nematic_tensors} for details. 
This definition provides a robust definition of reference frame 
if the direction of nematic order varies as function of spatial position.
Indeed, the visualization of nematic cell polarity in liver tissue
indicates a curved director field of nematic cell polarity on the lobule-level, 
see Fig.~\ref{fig:inertia_box}DE.

Below, we additionally consider a variation of this theme, 
where the tripod of reference axes $\n$, $\m$, $\l$ is not given by a local average, 
but by a second set of biaxial objects (namely the local anisotropy of the sinusoid transport network in the liver).

The most important difference between the traditional definition of the OOP, $S$, $P$, $D$, $C$, 
and our definition of COOP, $\coS$, $\coP$, $\coD$, $\coC$, 
is that the permutation $\pi$ of principal axes, and $\rho$ of reference axes
is determined by the orientational order of the ensemble itself for OOP, 
but prescribed by the eigenvalues of a nematic tensor for COOP.
This apparently small change in the mathematical definition
renders co-orientational order parameters (COOP) a robust metric that is applicable also in the case of curved director fields,
where classical order parameters (OOP) may change discontinuously.

Below, 
we will apply these biaxial co-orientational order parameters to quantify the alignment of nematic cell polarity of hepatocytes in liver tissue.

\paragraph{Geometric meaning of order parameters.}

\begin{figure}[!b] 
	\centering 
	\includegraphics[width=0.7\textwidth]{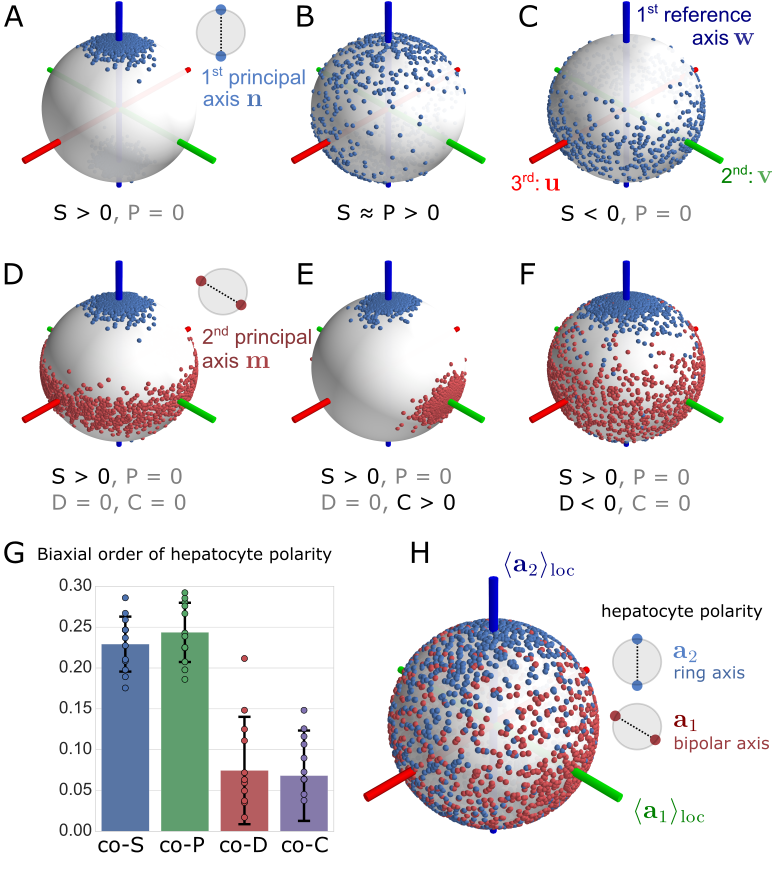}
	\caption{{\bf Four biaxial order parameters applied to liver tissue.}
	\newline
	(A)
Ensemble of first principal axes $\n$ 
that display \textit{prolate nematic order} with respect to the first reference axis $\w$, 
corresponding to $S>0$.
Each axis $\n$ of the ensemble is represented by an antipodal pair of blue points.
First reference axis $\w$ (blue line), second reference axis $\v$ (green), third reference axis $\u$ (red). 
(B) 
Example of a \textit{phase biaxial} distribution with nematic alignment towards the first reference axis $\w$
and strong anisotropic fluctuations biased towards the third reference axis $\u$ (red).
(C) 
Example of \textit{oblate nematic order} with respect to the first reference axis $\w$.
(D) 
Ensemble of tripods of principal axes $\n$, $\m$, $\l$ that displays \textit{prolate nematic order} 
of the first principal axis $\n$ (blue points) with respect to the first reference axis $\w$ (blue), 
but no additional order of the second principal axis $\m$ (red); third principal axis not shown.
(E) 
Example of \textit{molecular biaxial order} quantified by the order parameter $C$.
Here, the first principal axis $\n$ displays prolate nematic order as in panel D,
while the second principal axis $\m$ (red) is additionally biased towards the second reference axis $\v$ (green).
(F) 
A second type of molecular biaxial order is measured by the order parameter $D$.
Here, the first principal axis $\n$ (blue dots) exhibits nematic order with respect to the first reference axis $\w$ (blue).
Fluctuations of the second principal axis $\m$ (red dots) are also biased towards $\w$. 
(G) 
Co-orientational order parameters quantify biaxial order of hepatocytes in liver tissue 
(mean$\pm$s.d., $n=12$ tissue samples). 
The local reference system was chosen as a local average with punctured Gaussian kernel, see text for details. 
(H) 
Spherical distribution of apical ring axis $\a_2$ (blue dots) and apical bipolar axis $\a_1$ (red dots) of hepatocyte cell polarity
relative to the reference axes $\w=\langle\a_2\rangle_\mathrm{loc}$ (blue), $\v=\langle\a_1\rangle_\mathrm{loc}$ (green), $\u=\v\times\w$ (red),
illustrating the quantitative analysis in panel G.
}
\label{fig:scalar_order_parameter}
\end{figure}

We illustrate the geometric meaning of the orientational order parameters introduced in Eq.~(\ref{eq:SPDC}), 
see Fig.~\ref{fig:scalar_order_parameter}.
The case of co-orientational order parameters defined in Eq.~(\ref{eq:coSPDC}) is analogous
if principal axes $\n$, $\m$, $\l$ are plotted relative to the reference axes $\w$, $\v$, $\u$. 

When $S>0$ and all other order parameters vanish, as in panel \ref{fig:scalar_order_parameter}A, 
the ensemble is said to possess \textit{uniaxial prolate order}
(also called cluster-type order~\cite{fisher_1987}).
Such uniaxial orderings are axially symmetric around their first reference axis $\w$.
If fluctuations of the first principal axis $\n$ are anisotropic, 
as shown in panel \ref{fig:scalar_order_parameter}B, the ensemble is said to possess \textit{phase-biaxial order}.
This is quantified by the magnitude of the order parameter $P$.
In panel \ref{fig:scalar_order_parameter}C, an axially-symmetric distribution with $S<0$ is shown,
termed \textit{uniaxial oblate} order, 
where the first principal axis $\n$ scatters close to the `equator' 
(with north-pole south-pole axis set by $\w$).
This type of uniaxial order is occasionally called \textit{girdle order} \cite{fisher_1987}.

So far, we only examined the distribution of the first principal axis $\n$, 
which is quantified by the order parameters $S$ and $P$.
We now turn to the full description of biaxial nematic order, characterizing the distribution of a tripod of axes, $\n$, $\m$, $\l$.
In panels \ref{fig:scalar_order_parameter}D, E, and F, we show examples of an additional ordering of a second principal axis $\m$,
which are quantified by the other two order parameters $D$ and $C$.
We illustrate distributions of the second principal axis $\m$ by antipodal pairs of red points on the sphere.
Panel D shows the reference case of an uniaxial prolate distribution as in panel \ref{fig:scalar_order_parameter}A.
In absence of any additional ordering, the axis $\m$ 
displays uniaxial oblate order, as it is must be perpendicular to the first principal axis $\n$.
This example demonstrates that the type of order (prolate or oblate) 
crucially depends on which axis is chosen as the first principal axis.

We now consider the case of an additional ordering of the second principal axis $\m$.
In panel \ref{fig:scalar_order_parameter}E, $\m$ aligns towards the second reference axis $\v$ (green).
This breaks axial symmetry around $\w$ 
for the second principal axis $\m$ (red), but not for the first principal axis $\n$ (blue).
Correspondingly, the order parameter $P$ describing the phase biaxiality of the first principal axis $\n$ remains zero, 
but the \textit{molecular biaxiality parameter} $C$ becomes nonzero. 
This parameter thus describes the deviation from axial symmetry with respect to the first reference axis $\w$
of the ensemble of second principal axes $\m$.
In contrast, both the first and second principal axis, $\n$ and $\m$, compete for the same reference axis $\w$ in panel F.
Correspondingly, their respective distributions remain axially symmetric around $\w$.
In this case, both $P$ and $C$ are zero, 
yet the \textit{molecular ordering parameter} $D$ is non-zero.

\subsection*{Application to liver tissue}
\label{sec:application_liver}

We now apply the framework of biaxial order parameters to quantify lobule-level patterns of nematic cell polarity in mouse liver tissue.
We first examine the co-orientational order of the apical nematic polarity of hepatocytes with respect to its own local average.
As detailed in the preceding section,
we compare the nematic polarity axes of each individual hepatocyte (introduced in Fig.~\ref{fig:inertia_box}D)
with a local reference frame, given by a local average of the tensors $\matu A$ with a punctured Gaussian kernel 
(illustrated in Fig.~\ref{fig:inertia_box}F).
This provides reference axes
$\w=\langle\a_2\rangle_\mathrm{loc}$,
$\v=\langle\a_1\rangle_\mathrm{loc}$, and
$\u=\v\times\w$
at each hepatocyte position.
(Mathematically, $\langle\cdot\rangle_\mathrm{loc}$ is defined using nematic tensors, 
see SI text \nameref{sec:gaussian_average_nematic_tensors} for details.)
We choose the first principal axis $\vecu n$ to point in the direction of the ring axis $\vecu a_2$,
and the second principal axis $\vecu m$ to point in the direction of the bipolar axis $\vecu a_1$.

This choice uniquely specifies the four co-orientational order parameters, see Fig.~\ref{fig:scalar_order_parameter}G.
As additional illustration, 
we show the distribution of nematic cell polarity axes relative to its local reference system, see Fig.~\ref{fig:scalar_order_parameter}H.
We find that the ring axis $\a_2$ (blue dots) is clustered around the first reference axis $\w=\langle\a_2\rangle_\mathrm{loc}$.
Correspondingly, the scalar order parameter $\coS$ of uniaxial nematic order is larger than zero.
Additionally, we find a statistically significant phase biaxiality with $\coP>0$, 
revealing that fluctuations of the ring axis $\n=\a_2$ are biased away from the average bipolar axis $\v=\langle\a_1\rangle_\mathrm{loc}$.
This phase biaxiality is also visible in the distribution plot on the sphere in Fig.~\ref{fig:scalar_order_parameter}H.
The second principal axis $\m$ (bipolar axis $\a_1$, red dots) also exhibits a weak ordering
with a bias towards $\v$ (averaged bipolar axis $\langle\a_1\rangle_\mathrm{loc}$, green) and away from $\w$, 
reflected by positive values of $\coC$ and $\coD$, respectively.
Thus, 
using co-orientational order parameters that compare nematic axes with a local average (omitting the central cell), 
we can rigorously assess biaxial order even in the presence of curved director fields.

\subsection*{Co-alignment of nematic cell polarity and local anisotropy of blood transport network}

\begin{figure}[htp] 
\centering
\includegraphics[width=0.8\textwidth]{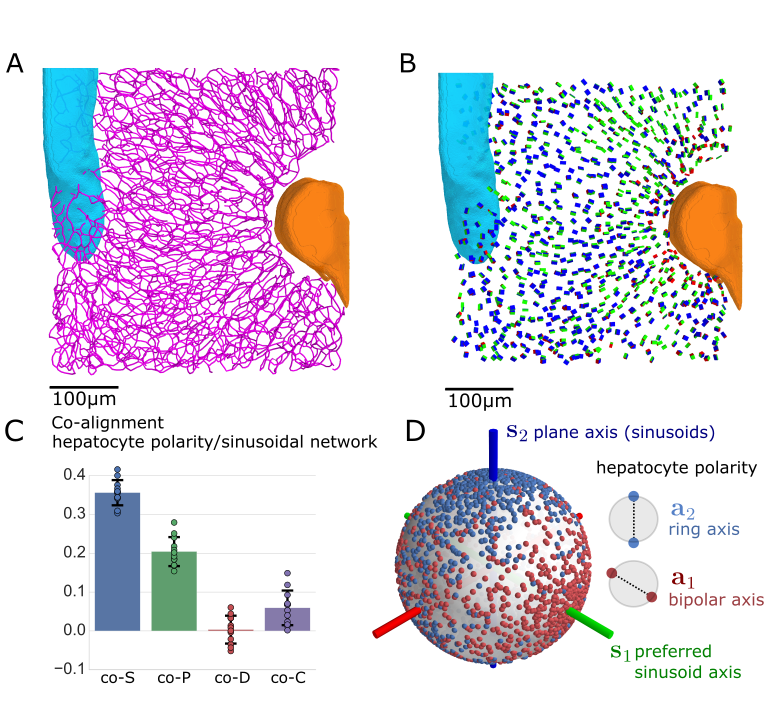}
\caption{{\bf Biaxial order of sinusoidal network correlates with nematic cell polarity.}
\newline
(A) 
Central lines of the sinusoidal network in the liver lobule
(same section of mouse liver tissue as in Fig.~\ref{fig:scalar_order_parameter};
central vein: cyan, portal vein: orange).
(B) 
The local anisotropy of the sinusoidal network is visualized by cuboids with equivalent moments-of-inertia tensor
(using spherical regions of interest centered at each hepatocyte position of $20\,\micron$ radius).
(C) 
Co-orientational order between apical nematic cell polarity and local anisotropy of the sinusoidal network
(mean$\pm$s.d., $n=12$ tissue samples). 
(D) 
Spherical distribution of apical ring axis (blue dots) and apical bipolar axis (red dots) in the reference frame of local sinusoidal network anisotropy.
}
\label{fig:apical_sinusoid_coop} 
\end{figure}

We can analyze nematic order of cell polarity not only within an ensemble of cells,
but also quantify the mutual alignment between cell polarity and auxiliary anisotropic structures such as transport networks.
As example, we analyze co-orientational order between apical nematic cell polarity of hepatocytes, and 
the local anisotropy of the blood-transporting sinusoidal network \cite{Morales2018,Karschau2019}.
Sinusoids are specialized blood vessels forming a network within the liver lobule \cite{Elias1955}.
Fig.~\ref{fig:apical_sinusoid_coop}A shows a central-line representation of the sinusoidal network.

We determine the local anisotropy of the sinusoidal network in the vicinity of each hepatocyte.
Specifically, if $\vecu{d}_k$ are unit vectors parallel to straight network segments, 
$\vecu x_{k}$ their midpoint positions and $l_{k}$ their respective lengths, we define nematic tensors $\matu{S}$ at each hepatocyte position $\vecu x^{(i)}$
\begin{align}
\label{eq:S}
\matu S (\vecu{x}^{(i)}) 
= 
\sum\limits_k 
w(\vecu{x}_k - \vecu{x}^{(i)}) \, l_k
\left( 
\vecu{d}_k \otimes \vecu{d}_k - \frac{1}{3} \mathbbm{1} 
\right)
\quad .
\end{align}
Here, $w(\x)$ is a weighting function normalized as $\sum_k w(\x_k-\x^{(i)}) l_k = 1$.
We choose $w(\x)$ as a binary cutoff with fixed radius of 20\,\textmu m around the center of each hepatocyte.
The geometric meaning of $\matu S$ can be understood as follows:
The eigenvector $\vecu s_1$, corresponding to the largest eigenvalue, characterizes the direction of preferred sinusoid orientation and will be referred to as \textit{preferred axis}.
The eigenvector $\vecu s_2$, corresponding to the smallest eigenvalue, defines the normal to a plane in which sinusoids orientations are preferentially distributed, 
and will be referred to as \textit{plane axis} in the following.
The biaxial anisotropy of the sinusoidal network with a distinguished plane axis 
is indicative of a local layered order, where $\s_2$ represents the normal vector of a stack of parallel layers, 
which characterizes this layered order.
Approximately, $\s_2$ is parallel to both 
the centerline of the portal vein, and
the centerline of the central vein.
Fig.~\ref{fig:apical_sinusoid_coop}B shows the spatial distribution of these nematic axes, using cuboids with equivalent moments of inertia.
The pattern of network anisotropy is similar to the averaged pattern of apical cell polarity.
Fig.~\ref{fig:apical_sinusoid_coop}C shows the four co-orientational order parameters between apical nematic cell polarity and local anisotropy of the sinusoidal network.
We find that the ring axis $\a_2$ of apical cell polarity is well-aligned with the plane axis $\s_2$ of the local sinusoid anisotropy.
For our choice of axes, this is quantified by the order parameter $\coS$. 
We also find phase-biaxiality of this axis, reflected by a non-zero value of $\coP$.
The other co-orientational order parameters  $\coD$ and $\coC$ are close to zero, 
i.e., we do not find a particular ordering of the bipolar cell polarity axis $\a_1$ relative to $\matu S$.  
The co-orientational order is also visualized as a spherical distribution plot in Fig.~\ref{fig:apical_sinusoid_coop}D,
highlighting the biaxial co-alignment between two different local anisotropies in liver tissue.

\subsection*{Minimal model for co-orientational order}

We present a minimal interaction model that can quantitatively reproduce the co-alignment between hepatocyte cell polarity and the biaxially anisotropic sinusoidal network.
If we account only for the ring axis $\a_2$ of hepatocytes,
the leading order term of an effective interaction energy is dictated by symmetry and reads
\begin{align}
H &= \lambda \, \left(\a_2 \otimes \a_2 \right) : \matu{S} \quad . 
\label{eq:hamiltonian_a3}
\end{align}
Here, $\matu{A} : \matu{E}$ denotes the contraction of two tensors $\matu{A}$ and $\matu{E}$.
Thus, we treat the hepatocytes as uniaxial objects, whereas we retain the biaxial anisotropy of the sinusoid network.
The choice of Eq.~(\ref{eq:hamiltonian_a3}) is motivated by a general Landau theory of liquid crystals, see \cite{matteis_2008,luckhurst_2015}.
We calculate the order parameters of an ensemble of axes according to the Boltzmann distribution following this Hamiltonian,
using inverse sampling. 
The control parameter $\lambda$ is measured in units of an effective temperature 
that mimics dynamic processes that reduce spatial order \cite{Cugliandolo:2011}.
We emphasize that we do not consider liver tissue to represent a thermodynamic equilibrium, 
despite our use of Eq.~(\ref{eq:hamiltonian_a3}). 
Instead, Eq.~(\ref{eq:hamiltonian_a3}) represents a phenomenological model that addresses a competition between dynamic processes that either generate or reduce spatial order, respectively.

\begin{figure}[htp] 
\centering
\includegraphics[width=0.7\textwidth]{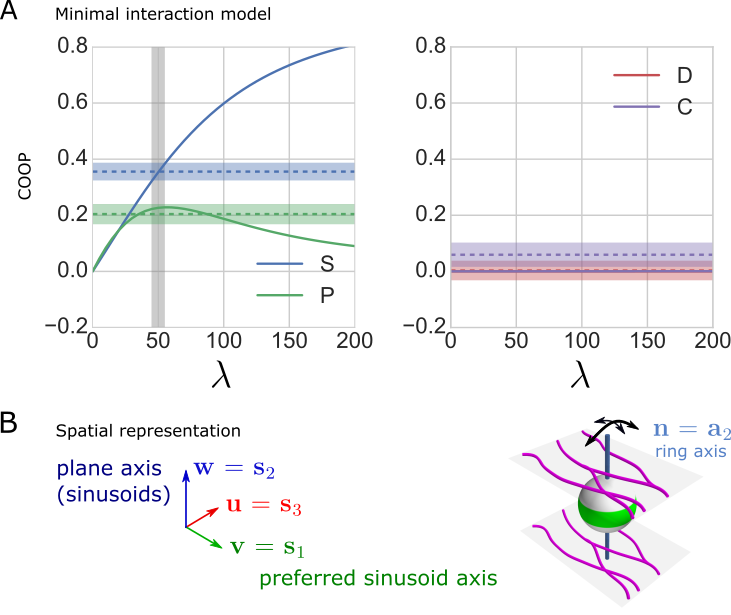}
\caption{{\bf Minimal interaction model reproduces biaxial order parameters for hepatocyte/sinusoid co-alignment.}
\newline
(A)
Simulated co-orientational order parameters (COOP) between
nematic cell polarity axes and local anisotropy of the sinusoidal network 
as function of the dimensionless interaction parameter $\lambda$ (solid lines),
see Eq.~(\ref{eq:hamiltonian_a3}).
Shaded regions indicate mean$\pm$s.d.\ of experimental values from Fig.~\ref{fig:apical_sinusoid_coop}C. 
The range of $\lambda$ for which all four order parameters agree in simulation and experiment is highlighted in gray. 
(B) 
Cartoon of hepatocyte/sinusoid co-alignment, 
where the ring axis $\a_2$ of hepatocyte polarity aligns parallel to the plane axis $\s_2$ of the local sinusoidal network. 
Fluctuations of $\a_2$ are biased away from the preferred axis $\s_1$ of the sinusoidal network. 
}
\label{fig:model1} 
\end{figure}

Figure~\ref{fig:model1}A displays computed COOPs,
together with the regions of order parameters found for the experimental data of liver tissue.
We find a range of values of the effective interaction parameter $\lambda$ (shaded gray in Fig.~\ref{fig:model1}A), 
where the minimal model adequately accounts for the experimental observed values of the co-orientational order parameters.
Thus, the interaction of the ring axis $\a_2$ of hepatocytes with the local anisotropy of the sinusoidal network 
is sufficient to account for the observed biaxial co-orientation.
Intriguingly, alternative models assuming either an interaction between $\matu S$ and the bipolar axis $\a_1$, 
or the full tensor $\matu A$, 
did not reproduce the observed co-orientational order, 
see SI text \nameref{sec:nematic_interaction_models}. 

This finding suggests the cartoon picture of sinusoid-hepatocyte co-alignment in liver tissue 
shown in Fig.~\ref{fig:model1}B.
We propose that the ring axis $\a_2$ of hepatocytes preferentially aligns parallel to the plane axis $\s_2$ of the local sinusoidal network.
Fluctuations of $\a_2$ break axial symmetry and are biased away from the preferred axis $\vecu s_1$ of the sinusoidal network.

\section*{Discussion} 

We presented a general method
to identify and quantify different types of cell polarity,
based on a multipole decomposition of surface patterns. 
We classify cell polarity as vectorial polarity, nematic polarity, or higher-order type.

We applied this method to three-dimensional reconstructions of epithelial tissue cells,
and the distribution of apical membrane markers on their surface \cite{morales-navarette_2015}.
We confirm that kidney cells predominantly display vectorial cell polarity. 
In contrast, hepatocytes from liver tissue are best characterized in terms of nematic cell polarity \cite{Morales2018}.
We propose a visualization method for spatial patterns of nematic cell polarity in terms of equivalent cuboids.
Applying this method to liver tissue reveals tissue-level patterns of coordinated cell polarity
that follows a curved director field on the level of a liver lobule \cite{Morales2018}.

To quantify this spatial order in a three-dimensional tissue,
we took inspiration from condensed matter physics.
Specifically, we generalized the four biaxial orientational order parameters (OOP) $S$, $P$, $D$, $C$ 
from the theory of liquid crystals \cite{maier_1959,gennes_1995,luckhurst_2015},
and generalized these as \textit{co-orientational order parameters}, which we then apply to quantify structural order in living matter.
Traditionally, OOP are used to quantify the partial alignment of anisotropic molecules, where each molecules characterized by a tripod of nematic axes
\cite{straley_1974,luckhurst_1980,carlsson_1991,berardi_2008}; here, we use COOP to quantify the partial alignment of nematic cell polarity axes in a tissue.

The co-orientational order parameters (COOP) introduced here have several advantages:
(i) unlike OOP, COOP do not depend on the choice of ordering of nematic axes and
(ii) change continuously if system parameters are smoothly varied,
yet are related to the classical OOP by simple linear transformations.
Moreover, 
(iii) COOP can be applied to curved director fields, and
(iv) be generalized to the case of an ensemble of pairs of biaxial objects in a straightforward manner.

Applying these COOP to mouse liver tissue, 
we show that the liquid-crystal order of nematic cell polarity of hepatocytes is biaxial,
which is a rare finding even in inanimate matter \cite{Mundoor2018}.
Furthermore, we found co-alignment between nematic cell polarity of hepatocytes and 
the local anisotropy of the sinusoidal network. 
This mutual alignment is not uniaxial, but of phase-biaxial type, i.e., its description requires two reference axes,
a preferred axis of the sinusoidal network (approximately parallel to the direction of blood flow \cite{Debbaut2012,Karschau2019}), 
and a plane axis, which characterizes local layered order of the sinusoidal network.
We conceptualized this biaxial order using a minimal interaction model, 
which quantitatively reproduces the COOP observed in the experimental data. 
The results from our minimal model provide insight into the previous observation of 
lobule-level spatial order of the bipolar axis ($\a_1$) of apical cell polarity of hepatocytes \cite{Morales2018}. 
In fact, our minimal model suggests that the observed order of the bipolar axes can be considered a consequence 
of the phase-biaxial co-alignment of the second cell polarity axis, the ring axis ($\a_2$), relative to the local anisotropy of the sinusoidal network. 

Our findings hint at a close interplay between hepatocyte polarity and the local anisotropy of the sinusoidal network.
A recent study involving some of the authors showed that interference with communication from sinusoids to hepatocytes 
disrupts the liquid-crystal order of both hepatocyte cell polarity and the sinusoidal network \cite{Morales2018}. 
Our analysis framework will allow to identify subtle changes in tissue architecture in the liver and other tissues
during development, genetic perturbations, or disease states.

\clearpage

\section*{Supporting Information}

\renewcommand{\theequation}{S\arabic{equation}}
\setcounter{section}{1} 
\setcounter{equation}{0}  
\renewcommand{\thefigure}{S\arabic{figure}}    
\setcounter{figure}{0}  
\renewcommand{\thetable}{S\arabic{table}}    
\setcounter{table}{0}  

\FloatBarrier
\paragraph{S1} 
\label{sec:symbols}
\textbf{List of symbols.}

Table~\ref{tab:symbols} contains a list of symbols used in the main text.

\begin{table}[htb]
\caption{
List of symbols used in the main text.
}
\label{tab:symbols}
\begin{adjustwidth}{0in}{0in}
\begin{tabular}{c|p{13cm}}
Symbol & Description \\
\hline
$\rho(\x)$ & scalar area density, e.g., density of polarity marker on cell surface \\[1mm]
$f(\x)$ & scalar area density on unit sphere $\mathcal{S}^2$; projection of $\rho(\x)$ \\[1mm]
$F_l(\x)$ & $l$-th mode of spherical Fourier transform of $f(\x)$, see Eq.~(\ref{eq:Y}) \\[1mm]
$f_l^m$ & $m$-th expansion coefficient of $F_l(\x)$, for expansion into spherical harmonics \\[1mm]
$\mathcal{S}^2$ & unit sphere \\[1mm]
$\int_{\mathcal{S}^2} \mathrm{d}^2\x$ & integral over unit sphere, using standard Euclidean measure \\[1mm]
$S_{f\!f}(l)$ & spherical power: $L^2$-norm of $l$-th mode $F_l(\x)$ of spherical Fourier transform of $f(\x)$ \\[3mm]
$\matu{A}$ & nematic tensor associated to surface density $f(\x)$, see Eq.~(\ref{eq:nematic_tensor}) \\[1mm]
$\alpha_1$, $\alpha_2$, $\alpha_3$ & eigenvalues of $\matu A$ \\[1mm]
$\a_1$, $\a_2$, $\a_3$ & 
eigenvectors of $\matu A$ corresponding to $\alpha_1$, $\alpha_2$, $\alpha_3$
with $\alpha_1\ge\alpha_3\ge\alpha_2$; 
we refer to $\a_1$ as \textit{bipolar axis} and $\a_2$ as \textit{ring axis} \\[1mm]
$\n$, $\m$, $\l$ & 
principal axes; the tripod of ortho-normal vectors
$\n$, $\m$, $\l$ represents a permutation of $\a_1$, $\a_2$, $\a_3$:
$\n=\a_{\pi(2)}$, $\m=\a_{\pi(3)}$, $\n=\a_{\pi(1)}$
with $\n$ first principal axis, $\m$ second principal axis, $\l$ third principal axis \\[1mm]
$\pi,\rho\in S_3$ & permutations of the indices $(1,2,3)$ \\[1mm]
$\matu Q$, $\matu B$ & traceless tensors, which characterize the tripod $\n$, $\m$, $\l$, see Eq.~(\ref{eq:QB}) \\[1mm]
$\matu{R}_Q$, $\matu{R}_B$ & rotation matrices that diagonalize $\matu Q$ and $\matu B$, respectively, see Eqs.~(\ref{eq:rq},\ref{eq:rb}) \\[1mm]
$\w$, $\v$, $\u$ &
reference axes; derived from either
a common eigenframe $\e_1$, $\e_2$, $\e_3$ of the tensors $\matu Q$ and $\matu B$, 
$\w=\e_{\rho(2)}$, $\v=\e_{\rho(3)}$, $\u=\e_{\rho(1)}$ (OOP), 
or from a second set of nematic tensors $\matu{E}$ 
with eigenvalues $\varepsilon_1$, $\varepsilon_2$, $\varepsilon_3$
that are ordered such that $\varepsilon_1\ge\varepsilon_3\ge\varepsilon_2$, 
and corresponding eigenvectors $\e_1$, $\e_2$, $\e_3$:
$\w=\e_2$, $\v=\e_3$, $\u=\e_1$ (COOP);
in both cases, we refer to the orthonormal vectors $\w$, $\v$, $\u$ as
$\w$ first reference axis, 
$\v$ second reference axis,
$\u$ third reference axis \\[1mm]
$S$, $P$, $D$, $C$ & 
orientational order parameters (OOP) that characterize biaxial order in an ensemble of tripods
$\a_1^{(i)}$, $\a_2^{(i)}$, $\a_3^{(i)}$ (with $D_{2h}$-symmetry), indexed by $i$; 
$S$ is the (uniaxial) nematic order parameter, 
$P$ is the phase-biaxial order parameter,
$D$ and $C$ are molecular biaxiality parameters that quantify order of a second nematic axis, see Eq.~(\ref{eq:SPDC}) \\[1mm]
$\coS$, $\coP$, $\coD$, $\coC$ & 
co-orientational order parameters (COOP), introduced here,
corresponding to a fixed ordering of principal axes,
$\n=\a_2$, $\m=\a_3$, $\l=\a_1$,
derived from the eigenvectors $\a_1$, $\a_2$, $\a_3$ of a nematic tensor $\matu A$ 
with corresponding eigenvalues $\alpha_1$, $\alpha_2$, $\alpha_3$ 
that are ordered such that $\alpha_1\ge\alpha_3\ge\alpha_2$,
and a fixed ordering of reference axes 
$\w=\e_2$, $\v=\e_3$, $\u=\e_1$ 
derived from the eigenvectors $\e_1$, $\e_3$, $\e_2$ of a second nematic tensor $\matu E$
with corresponding eigenvalues $\varepsilon_1$, $\varepsilon_2$, $\varepsilon_3$
that are ordered such that $\varepsilon_1\ge\varepsilon_3\ge\varepsilon_2$, see Eq.~(\ref{eq:coSPDC}) \\[3mm]
$\matu{S}$ & nematic tensor of local anisotropy of sinusoidal network, see Eq.~(\ref{eq:S}) \\[1mm]
$\s_1$, $\s_2$, $\s_3$ & eigenvectors of $\matu S$, 
corresponding to eigenvalues $\varepsilon_1$, $\varepsilon_2$, $\varepsilon_3$
with $\varepsilon_1\ge\varepsilon_3\ge\varepsilon_2$ \\[1mm]
$\mathbbm{1}$ & identity tensor with components $\mathbbm{1}_{\alpha\beta} = \delta_{\alpha\beta}$ \\[3mm]
$H$ & dimensionless Hamiltonian of minimal interaction model, see Eq.~(\ref{eq:hamiltonian_a3}) \\[1mm]
$\lambda$ & effective interaction parameter in $H$ \\[3mm]
$\matu I$ & moments-of-inertia tensor, see Eq.~(\ref{eq:I}) \\[1mm]
$a$, $b$, $c$ & 
side-lengths of equivalent cuboid for the visualization of nematic tensors $\matu A$, see SI text \nameref{sec:cuboid_visualization}; 
the convention $\alpha_1\ge\alpha_3\ge\alpha_2$ for the eigenvalues $\alpha_1$, $\alpha_2$, $\alpha_3$ of $\matu A$ implies $a\ge b\ge c$;
faces normal to edges of length $a$ are colored red, 
faces normal to edges of length $b$ are colored green, 
faces normal to edges of length $c$ are colored blue
\end{tabular}    
\end{adjustwidth}
\end{table}

\paragraph*{S2} 
\label{sec:spherical_projection}
\textbf{ Spherical projection of membrane protein density.}

\begin{figure}[htp] 
\centering
\includegraphics[width=0.5\textwidth]{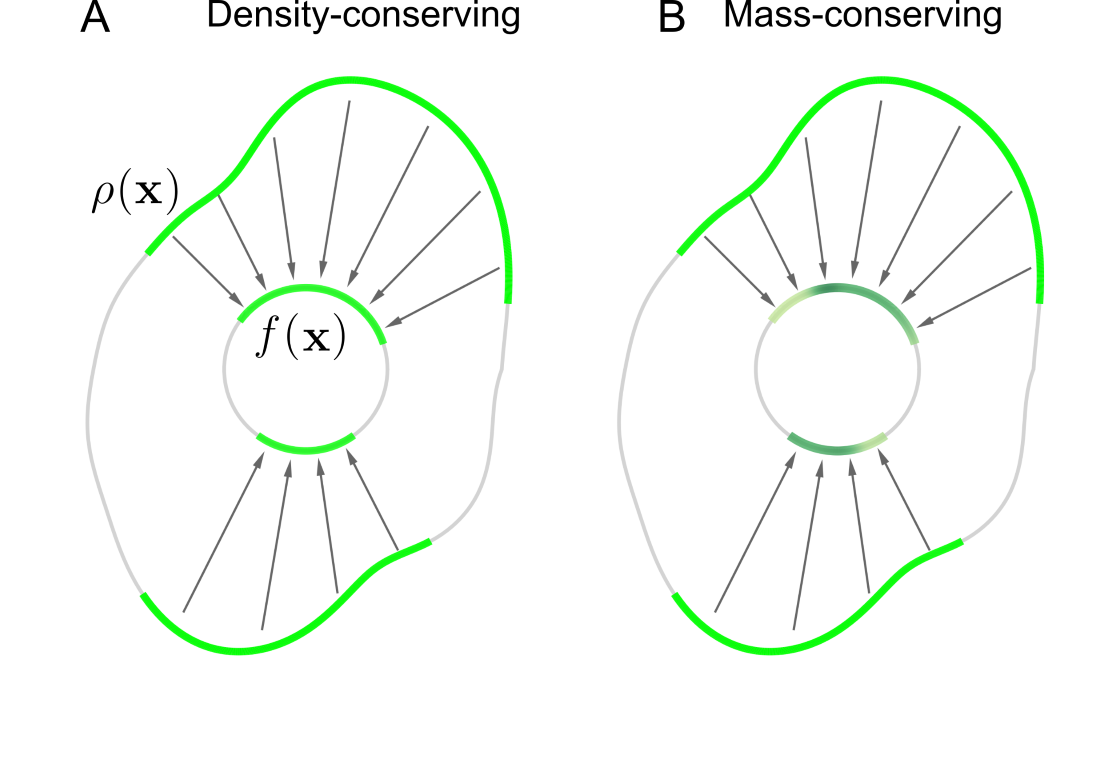}
\caption{
{\bf Schematic of spherical projection methods.}
\newline
We illustrate two methods to radially project a surface density (indicated in green) on a star-convex domain onto a co-centric sphere.
(A) 
In the variant used in the main text, the nominal value of the surface density is retained.
(B) 
Alternatively, one could multiply the local surface density by the relative change in area upon projection.
Thus, the total mass of the distribution is conserved.
However, the resultant spherical distribution will confound anisotropy of the original distribution and anisotropy of domain shape.
}
\label{fig:spherical_projection_2d} 
\end{figure}

We discuss two possible methods to project surface distributions $\rho(\x)$ on an arbitrary star-convex surface onto a sphere, 
shown schematically in Fig.~\ref{fig:spherical_projection_2d}.
The first method, depicted in panel A, retains the nominal value of the surface distribution.
In the second method, shown in panel B, the local surface density $\rho(\x)$ is weighted by the relative change in area upon projection.
In this case, the total mass of the distribution is preserved.
In the main text, we choose the first projection method because it ensures that a homogeneous distribution $\rho(\x)$ on the cell surface 
yields a projected distribution $f(\x)$ on the unit sphere that is again homogeneous.
By that, the effect of cell shape on the projected distribution is greatly reduced.
We confirmed that changing the projection method almost did not change computed cell polarity axes for most hepatocytes in liver tissue.

\paragraph{S3}
\label{sec:relation_second_mode}
\textbf{Relation between second mode spherical spectral power and order parameters.}

We can consider a distribution $p(\n)$ of nematic axes $\n$, represented by antipodal pairs of points,
also as a surface distribution on the unit sphere, which is symmetric with respect to a point reflection at the center.
Under this correspondence, 
the uniaxial order parameters $S$ and $P$ for $\n$ as first principal axis
are intimately linked to the expansion of $p(\n)$ into spherical harmonics, Eq.~(\ref{eq:Y}).
We assume that $p(\n)$ possesses $D_{2h}$-symmetry.
Without loss of generality, 
the reference axes $\u$, $\v$, $\w$ shall be aligned with the $x$, $y$, $z$ axis, respectively. 
Then, all coefficients $f_2^m$ 
of the second spherical mode $F_2(\x)=\sum_{m=-2}^2\, f_2^m Y_2^m(\x)$
that correspond to functions that are odd relative to a line reflection at the $z$-axis vanish,
i.e., $f_2^{-1}=f_2^1=0$ and $\mathrm{Im} f_2^{-2}=\mathrm{Im} f_2^2 = 0$.
Moreover, $f_2^2=f_2^{-2}$, since $p(\n)$ is real.

We thus find \cite{rosso_2007} 
\begin{align}
S = \sqrt{\frac{4\,\pi}{5}} f_2^0
\qquad 
P = \sqrt{6\,\frac{4\, \pi}{5}} f_2^2
\end{align}
Conversely, the spherical power in the second mode can be expressed in terms of $S$ and $P$
\begin{align}
S_{f\!f}(2) 
= \frac{1}{4\pi} \sum_{m=-2}^2 |f_2^m|^2 
= {\color{black} \frac{5}{(4\pi)^2} } \left( S^2 + \frac{1}{3} P^2 \right) \quad .
\end{align}

\paragraph{S4}
\label{sec:cuboid_visualization}
\textbf{Cuboid visualization of nematic cell polarity.}

In the main text, we present a method to visualize nematic tensors $\matu A$ by colored cuboids 
as shown in Fig.~\ref{fig:inertia_box}.
We provide additional details on this method.
For a surface density $f(\x)$ on the unit sphere $\mathcal{S}^2$,
the moments-of-inertia tensor $\matu{I}$ reads
\begin{align}
\label{eq:I}
\matu{I} = \int_{\mathcal{S}^2} \mathrm{d}^2\x \, \left( \mathbbm{1} - \x\otimes\x \right)\, \rho(\x),
\end{align}
i.e.,
\begin{align}
\matu{I} = \frac{2}{3} \left( F_0 \mathbbm{1} - \matu{A} \right)
\quad,
\label{eq:inertia_nematic}
\end{align}
where $\matu{A}$ is the nematic tensor associated to $f(\x)$, see Eq.~(\ref{eq:nematic_tensor}),
and $F_0=\int_{\mathcal{S}^2} \mathrm{d}^2\x\, \rho(\x)$.
Both tensors diagonalize in the same eigenframe.
The eigenvalues $\iota_1$, $\iota_2$, $\iota_3$ of $\matu{I}$ (called principal moments of inertia),
and the eigenvalues $\alpha_1$, $\alpha_2$, $\alpha_3$ of $\matu{A}$ are related by
$\iota_i = (2/3)F_0 - \alpha_i$, $i=1,2,3$ (for a suitable ordering of $\iota_i$).

In turn, the principal moments of inertia $\iota_i$ of a solid cuboid with side lengths $a$, $b$, $c$
are given by 
$\iota_1 = (b^2+c^2)/12$, 
$\iota_2 = (a^2+b^2)/12$, 
$\iota_3 = (a^2+c^2)/12$.
Using Eq.~\eqref{eq:inertia_nematic}, 
we find the side-lengths $a$, $b$, $c$ of a cuboid that has the same principal moments of inertia as $\matu{I}$ 
\begin{align}
a^2 &= 6 \left( \frac{2}{3}F_0 + \alpha_1 - \alpha_2 - \alpha_3 \right) \quad,  \\
b^2 &= 6 \left( \frac{2}{3}F_0 + \alpha_3 - \alpha_1 - \alpha_2 \right) \quad , \\ 
c^2 &= 6 \left( \frac{2}{3}F_0 + \alpha_2 - \alpha_3 - \alpha_1 \right) \quad . 
\end{align}
In plots, cuboids are rescaled by a constant factor.

\paragraph{S5}
\label{sec:gaussian_average_nematic_tensors}
\textbf{Gaussian average of nematic tensors.}

The coarse-grained orientation patterns shown in Fig.~\ref{fig:inertia_box}E are calculated 
from the nematic tensors $\matu A$ of individual hepatocytes by averaging with a Gaussian kernel.
Specifically, given nematic tensors $\matu{A}^{(i)}$ at cell center locations $\x^{(i)}$, 
the coarse-grained tensor $\langle \matu{A}\rangle_\mathrm{loc}(\x)$ at location $\x$ is calculated by
\begin{align}
\langle
\matu{A}
\rangle_\mathrm{loc}(\x) 
=
\sum_{i\neq j} 
\frac{1}{(2\pi\,\sigma^2)^{3/2}} 
\exp\left(
-\frac{ |\x^{(i)} - \x|^2}{2\,\sigma^2}
\right) 
\matu{A}^{(i)} \quad.
\label{eq:gauss_averaging}
\end{align}
Here, $\sigma$ denotes the standard deviation of the Gaussian kernel, which sets the length-scale of coarse-graining.
Note that we used a ``punctured'' Gaussian averaging kernel that omits the tensor of the central cell, 
thereby avoiding any bias.
As a side-node, instead of averaging nematic tensors $\matu{A}$, 
each derived from an individual surface distribution $f^{(i)}(\x)$, 
we could have equivalently averaged the surface distributions first, 
and then computed $\langle\matu{A}\rangle_\mathrm{loc}$ 
as the nematic tensor of an averaged surface distribution $\langle f(\x)\rangle_\mathrm{loc}$. 
{\color{black}
For the principal axes of the averaged tensor $\langle\matu{A}\rangle_\mathrm{loc}$, 
we write $\langle \a_1 \rangle_\mathrm{loc}$, $\langle \a_2 \rangle_\mathrm{loc}$, $\langle \a_3 \rangle_\mathrm{loc}$, 
for short.
}

\paragraph{S6}
\label{sec:nematic_interaction_models}
\textbf{Nematic interaction models.}

\begin{figure}[htp] 
    \centering
    \includegraphics[width=0.6\textwidth]{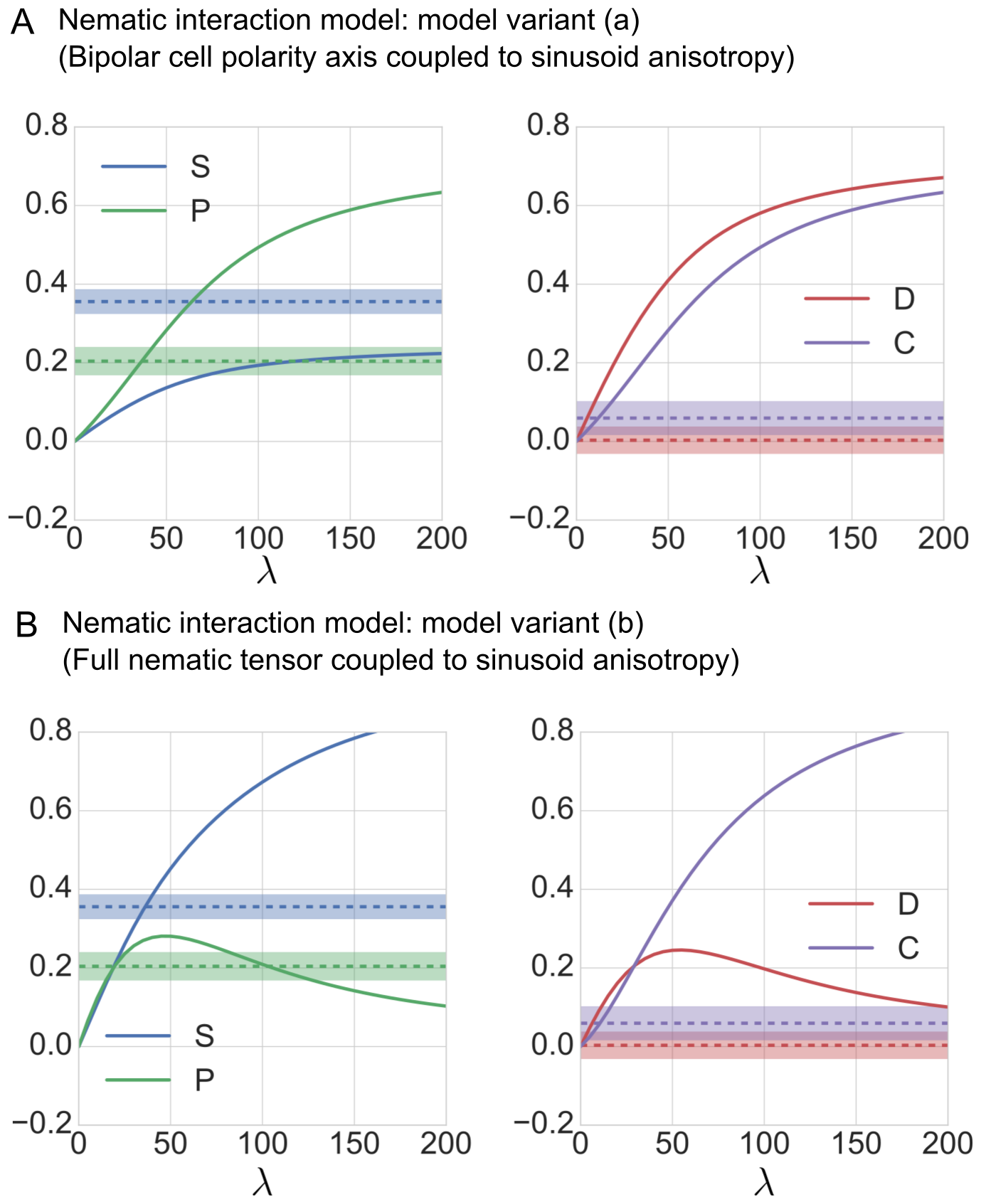}
    \vspace{1em}
    \caption{{\bf Alternative minimal interaction models ruled out by experimental data.}
(A)
Co-orientational order parameters predicted by a variant of the minimal interaction model, 
where only the bipolar apical nematic axis $\a_1$ of hepatocytes is coupled to the anisotropy of the local sinusoidal network, 
see Eq.~(\ref{eq:model2}).
There exists no value of the effective interaction parameter $\lambda$ 
for which simulation results are consistent with the experimental values 
(shaded region: mean$\pm$s.d., $n=12$ tissue samples).
(B)
Co-orientational order parameters predicted by a second variant of the minimal interaction model, see Eq.~(\ref{eq:model3}).
Here, the full apical nematic polarity tensor $\matu{A}$ of hepatocytes is coupled to the anisotropy of the local sinusoidal network.
Again, there exists no value of $\lambda$ consistent with the experimental data.
\newline}
\label{fig:som_figure2}
\end{figure}

In addition to the interaction proposed in Eq.~\eqref{eq:hamiltonian_a3}, 
two model variants are conceivable:
(a) the bipolar axis $\a_1$ of hepatocytes could be coupled to the local anisotropy tensor $\matu{S}$ of the sinusoidal network, or
(b) the full nematic tensor $\matu{A}$ of hepatocyte polarity, which comprises both the ring and the bipolar axes, could couple to $\matu{S}$, 
which correspond to respective effective interaction energies
\refstepcounter{equation}\label{eqn:model23}
\begin{align}
\tag{\theequation a}\label{eq:model2}
\text{model variant (a): }\ 
H &= \lambda\ \left(\a_1 \otimes \a_1\right) : {\matu S} \\
\tag{\theequation b}\label{eq:model3}
\text{model variant (b): }\ 
H &= \lambda\ {\matu A} : {\matu S} 
\end{align}
Fig.~\ref{fig:som_figure2} shows simulation results for these two alternative models.
We find that these alternative models cannot account 
for the experimentally observed values of the co-orientational order parameters.

\paragraph{S7} \textbf{Effect of axes permutations on orientational order parameters $S, P, D, C$}
\label{sec:order_parameter_permutations}

\begin{figure}[htp] 
\centering
\includegraphics[width=0.5\textwidth]{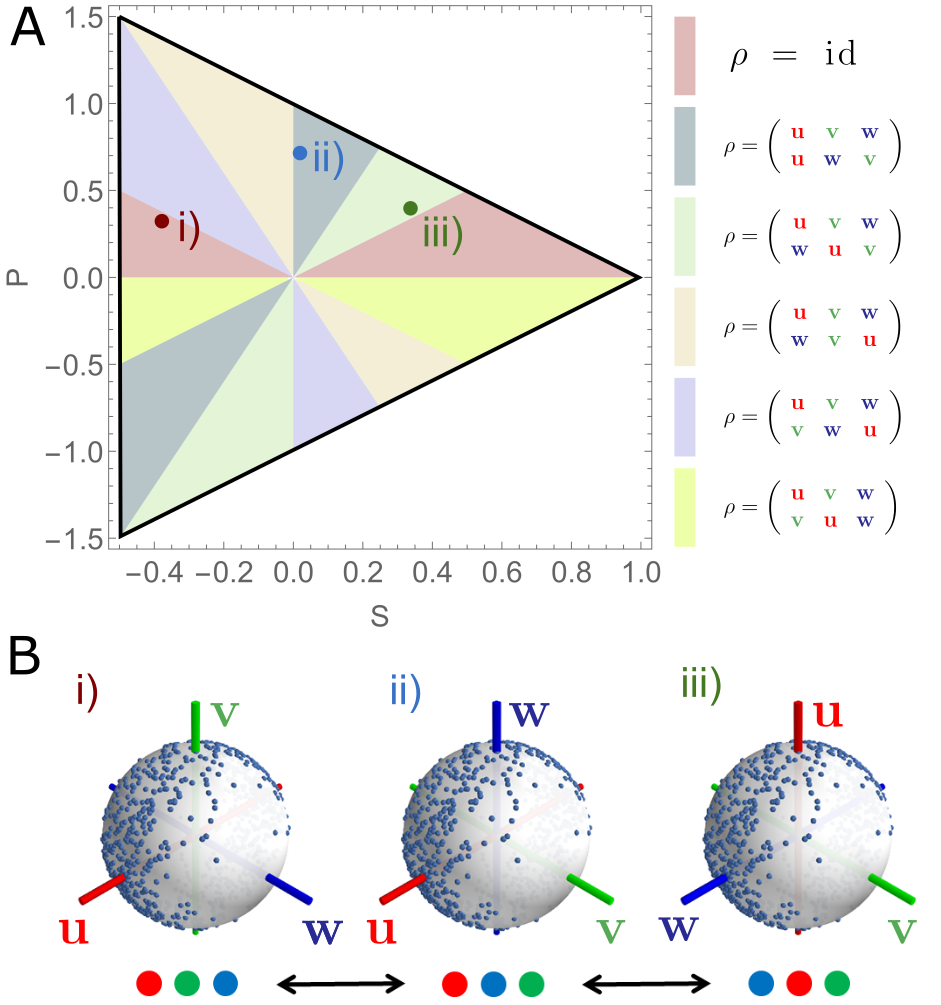}
\vspace{1em}
\caption{{\bf Action of axes permutation on orientational order parameters (OOP).}
(A)
Illustration of the action of the permutation group $G=S_3\times S_3$
(which permutes principal and reference axes, each) 
on the four-dimensional $(S,P,D,C)$-space.
Shown is the section defined by $D=C=0$,
corresponding to the subspace spanned by $S$ and $P$ that describes orientational order of a single axis. 
Colored regions show the tessellation of this subspace under the action of 
the permutation group $S_3$ of reference axes $\u$, $\v$, $\w$. 
The red region corresponds to a common convention in the theory of liquid crystals~\cite{rosso_2007, luckhurst_2015}.
(B)
Example distribution of first principal axis $\n$ (antipodal pairs of blue points),
displaying phase-biaxial order.
Three different permutations of the reference axes (cases i, ii, iii)
give rise to three different sets of order parameters for the same distribution of principal axes
(corresponding values of $S$ and $P$ indicated in panel A).
}
\label{fig:order_parameter_permutations}
\end{figure}

The orientational order parameters (OOP) $S$, $P$, $D$, $C$ defined in Eq.~(\ref{eq:SPDC})
change under a permutation $\pi\in S_3$ of the principal axes,
$\n=\a_{\pi(2)}$, $\m=\a_{\pi(1)}$, $\l=a_{\pi(3)}$,
as well as under a permutation $\rho\in S_3$ of the reference axes,
$\w=\e_{\rho(2)}$, $\v=\e_{\rho(1)}$, $\u=\e_{\rho(3)}$.
The action of the direct product of both permutation groups, $G = S_3 \times S_3$,
defines an equivalence relation on the four-dimensional space of 4-tuples $(S, P, D, C)$,
where each $G$-orbit defines one equivalence class that corresponds to the same state of orientational order.
Fig.~\ref{fig:order_parameter_permutations} illustrates the action of the permutation group $G$
on a two-dimensional section of the four-dimensional $(S, P, D, C)$-space.

\newsavebox{\rhomatalt}
\savebox{\rhomatalt}{
$\rho = 
\left( \begin{smallmatrix} 
\mathbf{u} & \mathbf{v} & \mathbf{w} \\ 
\mathbf{v} & \mathbf{u} & \mathbf{w}  
\end{smallmatrix} \right)$}
\newsavebox{\rhomat}
\savebox{\rhomat}{$\rho = \left( \begin{smallmatrix} 3 & 2 & 1 \\ 2 & 3 & 1 \end{smallmatrix} \right)$}

Table \ref{tab:spdc_permutation} lists the transformation 
of the orientational order parameters (OOP) $S$, $P$, $D$, $C$ under the action of the group $G$.

\begin{table}[htb]
\begin{adjustwidth}{0in}{0in}
    \begin{align*}
    \begin{array}{c|c|cccc}
    \pi     & \rho    & S & P & D & C \\
    \hline
    \hline
(\mathbf{l}\mathbf{m}\mathbf{n}) & (\mathbf{u}\mathbf{v}\mathbf{w}) & S & P & D & C \\[1mm]
 & (\mathbf{u}\mathbf{w}\mathbf{v}) & \frac{1}{2}(-S-P) & \frac{1}{2}(-3S+P) & \frac{1}{2}(-D-3C) & \frac{1}{2}(-D+C) \\[1mm]
 & (\mathbf{v}\mathbf{u}\mathbf{w}) & S & -P & D & -C \\[1mm]
 & (\mathbf{v}\mathbf{w}\mathbf{u}) & \frac{1}{2}(-S-P) & \frac{1}{2}(3S-P) & \frac{1}{2}(-D-3C) & \frac{1}{2}(D-C) \\[1mm]
 & (\mathbf{w}\mathbf{u}\mathbf{v}) & \frac{1}{2}(-S+P) & \frac{1}{2}(-3S-P) & \frac{1}{2}(-D+3C) & \frac{1}{2}(-D-C) \\[1mm]
 & (\mathbf{w}\mathbf{v}\mathbf{u}) & \frac{1}{2}(-S+P) & \frac{1}{2}(3S+P) & \frac{1}{2}(-D+3C) & \frac{1}{2}(D+C) \\[1mm]
\hline
(\mathbf{l}\mathbf{n}\mathbf{m}) & (\mathbf{u}\mathbf{v}\mathbf{w}) & \frac{1}{2}(-S-D) & \frac{1}{2}(-P-3C) & \frac{1}{2}(-3S+D) & \frac{1}{2}(-P+C) \\[1mm]
 & (\mathbf{u}\mathbf{w}\mathbf{v}) & \frac{1}{4}(S+P+D+3C) & \frac{1}{4}(3S-P+3D-3C) & \frac{1}{4}(3S+3P-D-3C) & \frac{1}{4}(3S-P-D+C) \\[1mm]
 & (\mathbf{v}\mathbf{u}\mathbf{w}) & \frac{1}{2}(-S-D) & \frac{1}{2}(P+3C) & \frac{1}{2}(-3S+D) & \frac{1}{2}(P-C) \\[1mm]
 & (\mathbf{v}\mathbf{w}\mathbf{u}) & \frac{1}{4}(S+P+D+3C) & \frac{1}{4}(-3S+P-3D+3C) & \frac{1}{4}(3S+3P-D-3C) & \frac{1}{4}(-3S+P+D-C) \\[1mm]
 & (\mathbf{w}\mathbf{u}\mathbf{v}) & \frac{1}{4}(S-P+D-3C) & \frac{1}{4}(3S+P+3D+3C) & \frac{1}{4}(3S-3P-D+3C) & \frac{1}{4}(3S+P-D-C) \\[1mm]
 & (\mathbf{w}\mathbf{v}\mathbf{u}) & \frac{1}{4}(S-P+D-3C) & \frac{1}{4}(-3S-P-3D-3C) & \frac{1}{4}(3S-3P-D+3C) & \frac{1}{4}(-3S-P+D+C) \\[1mm]
\hline
(\mathbf{m}\mathbf{l}\mathbf{n}) & (\mathbf{u}\mathbf{v}\mathbf{w}) & S & P & -D & -C \\[1mm]
 & (\mathbf{u}\mathbf{w}\mathbf{v}) & \frac{1}{2}(-S-P) & \frac{1}{2}(-3S+P) & \frac{1}{2}(D+3C) & \frac{1}{2}(D-C) \\[1mm]
 & (\mathbf{v}\mathbf{u}\mathbf{w}) & S & -P & -D & C \\[1mm]
 & (\mathbf{v}\mathbf{w}\mathbf{u}) & \frac{1}{2}(-S-P) & \frac{1}{2}(3S-P) & \frac{1}{2}(D+3C) & \frac{1}{2}(-D+C) \\[1mm]
 & (\mathbf{w}\mathbf{u}\mathbf{v}) & \frac{1}{2}(-S+P) & \frac{1}{2}(-3S-P) & \frac{1}{2}(D-3C) & \frac{1}{2}(D+C) \\[1mm]
 & (\mathbf{w}\mathbf{v}\mathbf{u}) & \frac{1}{2}(-S+P) & \frac{1}{2}(3S+P) & \frac{1}{2}(D-3C) & \frac{1}{2}(-D-C) \\[1mm]
\hline
(\mathbf{m}\mathbf{n}\mathbf{l}) & (\mathbf{u}\mathbf{v}\mathbf{w}) & \frac{1}{2}(-S-D) & \frac{1}{2}(-P-3C) & \frac{1}{2}(3S-D) & \frac{1}{2}(P-C) \\[1mm]
 & (\mathbf{u}\mathbf{w}\mathbf{v}) & \frac{1}{4}(S+P+D+3C) & \frac{1}{4}(3S-P+3D-3C) & \frac{1}{4}(-3S-3P+D+3C) & \frac{1}{4}(-3S+P+D-C) \\[1mm]
 & (\mathbf{v}\mathbf{u}\mathbf{w}) & \frac{1}{2}(-S-D) & \frac{1}{2}(P+3C) & \frac{1}{2}(3S-D) & \frac{1}{2}(-P+C) \\[1mm]
 & (\mathbf{v}\mathbf{w}\mathbf{u}) & \frac{1}{4}(S+P+D+3C) & \frac{1}{4}(-3S+P-3D+3C) & \frac{1}{4}(-3S-3P+D+3C) & \frac{1}{4}(3S-P-D+C) \\[1mm]
 & (\mathbf{w}\mathbf{u}\mathbf{v}) & \frac{1}{4}(S-P+D-3C) & \frac{1}{4}(3S+P+3D+3C) & \frac{1}{4}(-3S+3P+D-3C) & \frac{1}{4}(-3S-P+D+C) \\[1mm]
 & (\mathbf{w}\mathbf{v}\mathbf{u}) & \frac{1}{4}(S-P+D-3C) & \frac{1}{4}(-3S-P-3D-3C) & \frac{1}{4}(-3S+3P+D-3C) & \frac{1}{4}(3S+P-D-C) \\[1mm]
\hline
(\mathbf{n}\mathbf{l}\mathbf{m}) & (\mathbf{u}\mathbf{v}\mathbf{w}) & \frac{1}{2}(-S+D) & \frac{1}{2}(-P+3C) & \frac{1}{2}(-3S-D) & \frac{1}{2}(-P-C) \\[1mm]
 & (\mathbf{u}\mathbf{w}\mathbf{v}) & \frac{1}{4}(S+P-D-3C) & \frac{1}{4}(3S-P-3D+3C) & \frac{1}{4}(3S+3P+D+3C) & \frac{1}{4}(3S-P+D-C) \\[1mm]
 & (\mathbf{v}\mathbf{u}\mathbf{w}) & \frac{1}{2}(-S+D) & \frac{1}{2}(P-3C) & \frac{1}{2}(-3S-D) & \frac{1}{2}(P+C) \\[1mm]
 & (\mathbf{v}\mathbf{w}\mathbf{u}) & \frac{1}{4}(S+P-D-3C) & \frac{1}{4}(-3S+P+3D-3C) & \frac{1}{4}(3S+3P+D+3C) & \frac{1}{4}(-3S+P-D+C) \\[1mm]
 & (\mathbf{w}\mathbf{u}\mathbf{v}) & \frac{1}{4}(S-P-D+3C) & \frac{1}{4}(3S+P-3D-3C) & \frac{1}{4}(3S-3P+D-3C) & \frac{1}{4}(3S+P+D+C) \\[1mm]
 & (\mathbf{w}\mathbf{v}\mathbf{u}) & \frac{1}{4}(S-P-D+3C) & \frac{1}{4}(-3S-P+3D+3C) & \frac{1}{4}(3S-3P+D-3C) & \frac{1}{4}(-3S-P-D-C) \\[1mm]
\hline
(\mathbf{n}\mathbf{m}\mathbf{l}) & (\mathbf{u}\mathbf{v}\mathbf{w}) & \frac{1}{2}(-S+D) & \frac{1}{2}(-P+3C) & \frac{1}{2}(3S+D) & \frac{1}{2}(P+C) \\[1mm]
 & (\mathbf{u}\mathbf{w}\mathbf{v}) & \frac{1}{4}(S+P-D-3C) & \frac{1}{4}(3S-P-3D+3C) & \frac{1}{4}(-3S-3P-D-3C) & \frac{1}{4}(-3S+P-D+C) \\[1mm]
 & (\mathbf{v}\mathbf{u}\mathbf{w}) & \frac{1}{2}(-S+D) & \frac{1}{2}(P-3C) & \frac{1}{2}(3S+D) & \frac{1}{2}(-P-C) \\[1mm]
 & (\mathbf{v}\mathbf{w}\mathbf{u}) & \frac{1}{4}(S+P-D-3C) & \frac{1}{4}(-3S+P+3D-3C) & \frac{1}{4}(-3S-3P-D-3C) & \frac{1}{4}(3S-P+D-C) \\[1mm]
 & (\mathbf{w}\mathbf{u}\mathbf{v}) & \frac{1}{4}(S-P-D+3C) & \frac{1}{4}(3S+P-3D-3C) & \frac{1}{4}(-3S+3P-D+3C) & \frac{1}{4}(-3S-P-D-C) \\[1mm]
 & (\mathbf{w}\mathbf{v}\mathbf{u}) & \frac{1}{4}(S-P-D+3C) & \frac{1}{4}(-3S-P+3D+3C) & \frac{1}{4}(-3S+3P-D+3C) & \frac{1}{4}(3S+P+D+C) \\[1mm]
\hline
\end{array}
\end{align*}
\end{adjustwidth}
\caption{
Transformation of the orientational order parameters $S$, $P$, $D$, $C$, 
under a permutation $\pi\in S_3$ of the principal axes, or
permutation $\rho\in S_3$ of the reference axes.
Permutations are shown in one-line notation (i.e., second row of Cauchy's two-line notation).
}
\label{tab:spdc_permutation}
\end{table}

\FloatBarrier
\paragraph{S8}
\label{sec:relation_biaxial_order_invariants}
\textbf{Relation between biaxial order parameters and invariants of moment tensors.}

We quantified orientational order of nematic tensors by the four classical order parameters $S$, $P$, $D$, $C$, 
as well as by co-orientational order parameters $\coS$, $\coP$, $\coD$, $\coC$.
We present yet a third variant to quantify biaxial order: invariants of moment tensors \cite{Scholich2018}.

We consider the first two moments, $\matu{T}$ and $\matu{V}$, of a distribution of nematic tensors $\matu{A}$
\begin{align}
\text{first moment: } && 
\matu{T} &= \langle\matu{A}\rangle \quad, 
\nonumber\\
\text{second moment:} &&
\matu{V} &= \langle \matu{A}:\matu{A} \rangle 
\text{ with components }
V_{\alpha\beta} = \angleave{A_{\alpha\gamma} A_{\gamma\beta}} \quad .
\end{align}
From these averaged tensors, we obtain scalar invariants by tensor contraction
\begin{align}
I_1 &= \mathrm{tr}\,\matu{A} ,\quad &
I_2 &= \mathrm{tr}\,\matu{A}^2 ,\quad &
I_3 &= \mathrm{tr}\,\matu{A}^3 ,\quad \nonumber\\
I_4 &= \mathrm{tr}\,\matu{V} ,\quad &
I_5 &= \mathrm{tr}\,\matu{V}^2 ,\quad &
I_6 &= \mathrm{tr}\,\matu{V}^3 ,\quad \ldots \quad .
\end{align}
Note that since $\matu{A}$ is traceless, 
all non-zero contractions of the rank-4 super-tensor $\langle A_{\alpha\gamma}A_{\delta\beta}\rangle$
can already be derived from the rank-2 tensor $\matu{V}$.

If the ensemble of tensors $\matu{A}^{(i)}$ exhibits $D_{2h}$-symmetry, 
the moment tensors $\matu{T}$ and $\matu{V}$ diagonalize in a common eigenframe~\cite{matteis_2009}.
The invariants $I_1,\ldots,I_6$ can then be expressed in terms of symmetric polynomials in the eigenvalues of these tensors.
Specifically, we denote the eigenvalues of $\matu{T}$ by $\mu_1$, $\mu_2$ and $\mu_3$, 
and the eigenvalues of $\matu{V}$ by $\nu_1$, $\nu_2$, $\nu_3$.
Then the invariants of tensor moments are given as
\begin{align}
I_1 &= \sum_{i=1}^3 \mu_{i} = 0
,\quad &
I_2 &= \sum_{i=1}^3 \mu_{i}^2 
,\quad &
I_3 &= \sum_{i=1}^3 \mu_{i}^3 
, 
\nonumber\\ 
I_4 &= \sum_{i=1}^3 \nu_{i}
,\quad &
I_5  &= \sum_{i=1}^3 \nu_{i}^2
, \quad &
I_6 &= \sum_{i=1}^3 \nu_{i}^3 
\quad , \ \dots \ .
\label{eq:invariants_eigenvalues}
\end{align}
Conversely, 
given the invariants $I_1$, $I_2$, $\dots$, $I_6$, 
we can compute the eigenvalues $\mu_1$, $\mu_2$, $\mu_3$, and  $\nu_1$, $\nu_2$, $\nu_3$,
yet only up to a permutation, 
by solving the polynomial system of equations, Eq.~\eqref{eq:invariants_eigenvalues}.

We now show how the tensor invariants are related to the classical order parameters $S$, $P$, $D$, $C$.
We make the simplifying assumption that all nematic tensors $\matu{A}$ of the ensemble
have identical eigenvalues
$\alpha_1$, $\alpha_2$, $\alpha_3$. 
As usual, we assume $\alpha_1\ge \alpha_3\ge\alpha_2$,
and denote the corresponding eigenvectors as $\n=\a_2$, $\m=\a_1$, $\l=\a_3$.
By Eq.~(\ref{eq:QB}), each tensor $\matu{A}$ is now associated with tensors $\matu{Q}$ and $\matu{B}$. 
We can write
\begin{equation}
\matu{A} = \xi_0\,\matu{Q} + \xi_1\,\matu{B}\quad,
\end{equation}
with weights $\xi_0$, $\xi_1$ that satisfy
$\alpha_2=\xi_0$, $\alpha_1=-(\xi_0+3\xi_1)/2$, $\alpha_3=-(\xi_0-3\xi_1)/2$.
Note that a permutation of eigenvalues also changes the weights $\xi_0$ and $\xi_1$.
Our usual ordering of eigenvalues with $\alpha_1\ge \alpha_3\ge \alpha_2$ corresponds to $0\ge \xi_0/3\ge \xi_1$.
In the general case, where the eigenvalues of $\matu{A}$ vary within the ensemble,
the invariants $I_j$ will depend on both the orientational order of the principal axes of $\matu{A}$,
as well as on the distribution of weights.

In the case of constant weights $\xi_0$, $\xi_1$, 
it follows 
$\matu{T} 
= \langle \matu{A}\rangle 
= \xi_0\,\langle\matu{Q}\rangle + \xi_1\,\langle\matu{B}\rangle$.
Likewise, the second moment $\matu{V}$ can be expressed as a linear superposition of
$\langle\matu{Q}\rangle$, $\langle\matu{B}\rangle$, and $\mathbbm{1}$ as 
\begin{equation}
\matu{V} = \zeta_0\, \langle\matu{Q}\rangle + \zeta_1\, \langle\matu{B}\rangle + \zeta_c\, \mathbbm{1}\quad,
\end{equation}
where 
$\zeta_0 = (\xi_0^2-3\xi_1^2)/2$, 
$\zeta_1 = - \xi_0\xi_1$,
$\zeta_c = (\xi_0^2+3\xi_1^2)/2$. 

We thus have a direct correspondence between the eigenvalues of the moment tensors 
and the orientational order parameters $S$, $P$, $D$, $C$ 
\begin{equation}
\begin{array}{ccccccc}
2\,\mu_1 &=& & -\xi_0 S & +\xi_0 P & \phantom{2}-\xi_1 D & +3\,\xi_1 C \ ,\\
2\,\mu_2 &=& & -\xi_0 S & -\xi_0 P & \phantom{2}-\xi_1 D & -3\,\xi_1 C \ ,\\
2\,\mu_3 &=& & 2\xi_0 S &          & +2\xi_1 D  & \ ,\\[1mm]
2\,\nu_1 &=& {\color{black} 2} \zeta_c & \phantom{2}-\zeta_0 S & +\zeta_0 P & \phantom{2}-\zeta_1 D & +3\,\zeta_1 C \ ,\\
2\,\nu_2 &=& {\color{black} 2} \zeta_c & \phantom{2}-\zeta_0 S & -\zeta_0 P & \phantom{2}-\zeta_1 D & -3\,\zeta_1 C \ ,\\
2\,\nu_3 &=& {\color{black} 2} \zeta_c & +2\zeta_0 S &          & +2\zeta_1 D & \quad .
\end{array}
\label{eq:eigenvalues_order_parameters}
\end{equation}
Note $\mu_1+\mu_2+\mu_3=0$, while $\nu_1+\nu_2+\nu_3=3\zeta_c$.

Together, 
Eq.~(\ref{eq:invariants_eigenvalues}) and Eq.~(\ref{eq:eigenvalues_order_parameters}) 
allow to compute the orientational order parameters $S$, $P$, $D$, $C$
from the invariants of tensor moments.
Note that the first tensor moment is not sufficient to determine the OOP, 
but that at least the second moment is needed. 
(In the non-generic case
{\color{black}
$\xi_0=-1$ and $\xi_1=\pm 1$ for which $\zeta_0 = \xi_0$ and $\zeta_1 = \xi_1$, 
}
also a third tensor moment needs to be taken into account.)

\begin{figure}[htp] 
\centering
\includegraphics[width=0.7\textwidth]{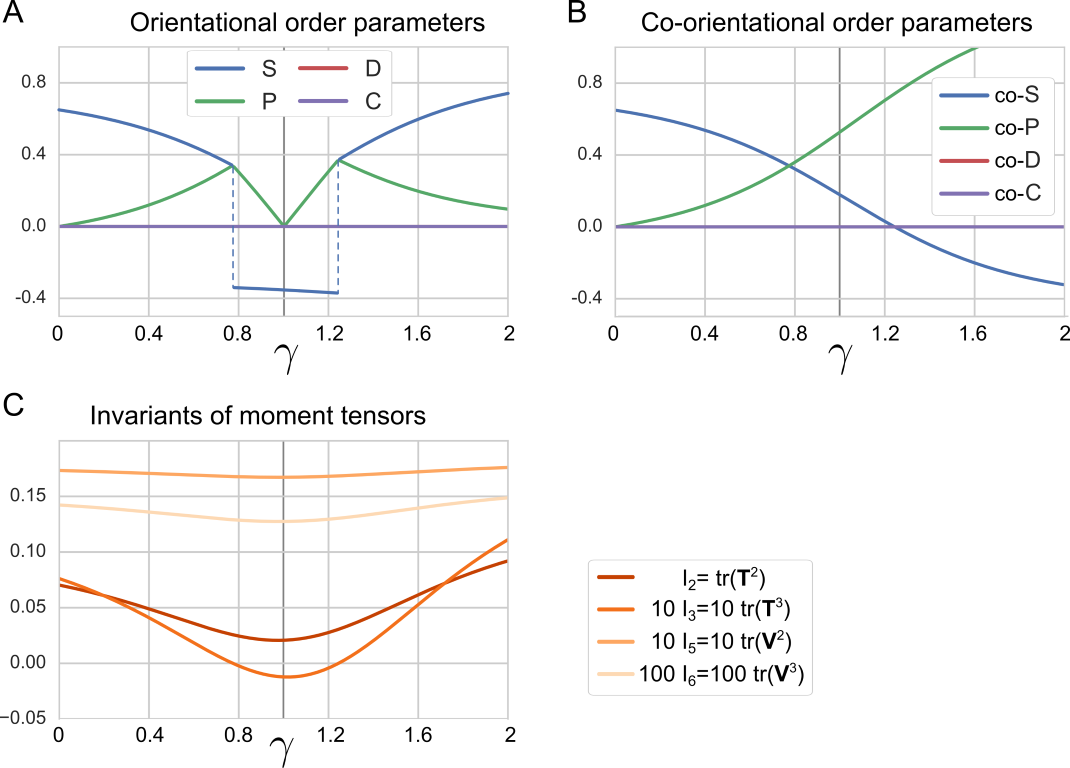}
\vspace{1em}
\caption{
{\bf 
Comparison of orientational order parameters (OOP), 
co-orientational order parameters (COOP), 
and invariants of moment tensors for phase-biaxial order.}
(A)
Orientational order parameters $S$, $P$, $D$, $C$
for a Boltzmann distribution $p\sim\exp(-H)$ of biaxial objects 
governed by the dimensionless Hamiltonian
{\color{black}
$H = -(\a_2\cdot\e_2)^2 -\gamma (\a_2\cdot\e_1)^2$ 
}
as function of the effective interaction parameter $\gamma$. 
The ordering $\pi$ of principal axes with $\n=\a_{\pi(2)}$ as well as
the ordering $\rho$ of reference axes with
$\w=\e_{\rho(2)}$, $\v=\e_{\rho(3)}$, $\u=\e_{\rho(1)}$
in the definition of $S$, $P$, $D$, $C$, Eq.~(\ref{eq:SPDC})
is chosen such that $|S|$ is maximal and $P\ge 0$, $C\ge 0$
(as common in the field of liquid crystals \cite{luckhurst_2015}).
Note the discontinuous change of $S$ and $P$ 
caused by a change in $\rho$.
(B) 
Same as panel A for the co-orientational order parameters $\coS$, $\coP$, $\coD$, $\coC$,
{\color{black}
where 
a fixed ordering $\pi =\mathrm{id}$ of principal axes and
a fixed ordering $\rho=\mathrm{id}$ of reference axes is used.
For this choice, 
Eq.~(\ref{eq:zannoni}) holds for $\gamma=0$, but not for general $\gamma$.
}
(C) 
Invariants of tensor moments as defined in Eq.~(\ref{eq:invariants_eigenvalues}) for the same system.
}
\label{fig:compare_spdc_tensor_invariants}
\end{figure}

We emphasize that the values of the invariants $I_1$, $I_2$, $\dots$, $I_6$ 
are independent of any ordering of axes, 
whereas the values of the orientational order parameters $S, P, D, C$ depend 
on the ordering of both the principal and the reference axes. 
This is reflected in 
Eq.~(\ref{eq:invariants_eigenvalues})
by the fact that $I_1, I_2, \dots, I_6$ do not change under 
neither a permutation of the eigenvalues $\mu_1$, $\mu_2$, $\mu_3$,
nor a permutation of the eigenvalues $\nu_1$, $\nu_2$, $\nu_3$. 
In contrast, 
Eq.~(\ref{eq:eigenvalues_order_parameters}) 
shows that $S$, $P$, $D$, $C$ 
depend on the ordering of eigenvalues.
As a consequence, the orientational order parameters $S$, $P$, $D$, $C$ can change discontinuously 
if system parameters are smoothly varied, while the invariants $I_1$, $I_2$, $\dots$, $I_6$ do not, 
see Fig.~\ref{fig:compare_spdc_tensor_invariants}.
Despite this desirable property of the invariants $I_1$, $I_2$, $\dots$, $I_6$,
the invariants lack the intuitive geometric interpretation of the orientational order parameters $S$, $P$, $D$, $C$.
The co-orientational order parameters $\coS$, $\coP$, $\coD$, $\coC$ introduced in the main text
combine the advantageous property of a smooth dependence on system parameters
with intuitive geometric interpretation. 

\
\clearpage
\section*{Acknowledgments}

We thank Samuel Safran for stimulating discussions.
AS acknowledges support from the Max-Planck society. 
Parts of this work were supported by the German Federal Ministry of Education and Research (BMBF) 
within the research network Systems Medicine of the Liver (LiSyM) [grant 031L0033 to LB, grant 031L0038 to MZ].
Parts of this work were supported by the European Research Council (ERC) (grant number 695646 to MZ).
BMF, FJ and MZ acknowledge support by the DFG through the Excellence Initiative 
by the German Federal and State Governments (Clusters of Excellence cfaed EXC 1056 and PoL EXC 2068). 

\nolinenumbers

\end{document}